\documentclass{article}

\usepackage{arxiv}

\usepackage{amsmath}
\usepackage{comment}
\usepackage{amssymb}
\usepackage[utf8]{inputenc} 
\usepackage{graphicx}
\usepackage{subcaption}
\usepackage[T1]{fontenc}    
\usepackage{url}            
\usepackage{booktabs}       
\DeclareMathOperator*{\argmax}{argmax}
\DeclareMathOperator*{\argmin}{argmin}
\usepackage{algorithm}
\usepackage{algorithmic}
\usepackage{amsfonts}       
\usepackage{nicefrac}       
\usepackage{microtype}      
\usepackage{lipsum}
\usepackage{hyperref}
\newtheorem{theorem}{Theorem}[section]

\newtheorem{lemma}[theorem]{Lemma}
\hypersetup{
    colorlinks=true,
    linkcolor=blue,
    filecolor=magenta,      
    urlcolor=cyan,
}

\title{Estimation of dynamical systems in noisy conditions and with constraints}

\author{
  Krishan Mohan Nagpal  
  \thanks{Krishan Nagpal is a Managing Director in Corporate Risk in Wells Fargo \& Co. The opinions expressed here are those of the author and do not represent those of his employer Wells Fargo \& Co.}\\
  Wells Fargo \& Co. \\
  \texttt{krishan$\_$nagpal@yahoo.com} \\
}

\begin{document}
\maketitle

\begin{abstract}
When measurements from dynamical systems are noisy, it is useful to have estimation algorithms that have low sensitivity to measurement noises and outliers. In the first set of results described in this paper we obtain optimal estimators for linear dynamical systems with $\epsilon$ insensitive loss functions. The  $\epsilon$ insensitive loss function, which is often used in Support Vector Machines, provides more stable and smooth estimates when the measurements are biased and very noisy as the algorithm tolerates small errors in prediction which in turn makes the estimates less sensitive to measurement noises. Apart from $\epsilon$ insensitive quadratic loss function, estimation algorithms are also derived for $\epsilon$ insensitive Huber M loss function which provides good performance in presence of both small noises as well as outliers. The advantage of Huber cost function based estimator in presence of outliers is due to the fact the error penalty function switches from quadratic to linear for errors beyond a certain threshold.  

The second set of results in the paper describe algorithms for estimation when apart from general description of dynamics of the system, one also has additional information about states and exogenous signals such as known range of some states or prior information about the maximum magnitude of noises/disturbances. While the proposed approaches have similarities to optimal $\mathcal{H}_2$ smoothing algorithm, the algorithms are not linear in measurements but are easily implemented as optimal estimates are obtained by solving a standard quadratic optimization problem with linear constraints. For all cases, algorithms are proposed not only for filtering and smoothing but also for prediction of future states. 
\end{abstract}

\section{Introduction}

The problems of prediction, filtering and smoothing involve estimating states of a dynamical system based on noisy measurements of the system. In prediction the goal is to predict state of the system at a future time without knowledge of the future measurements. In filtering, the goal is to estimate the current state of the system using all the past measurements while in smoothing one estimates both the current and past states of the system. When the dynamical system is linear and known, Kalman Bucy estimation framework, also sometimes described as $\mathcal{H}_2$ framework, provides optimal way to estimate states of the system when the power spectrum density of the noises and disturbances is known (\cite{anderson} and \cite{meditch}). Kalman-Bucy framework also has certain optimal worst case properties. For example Krener (\cite{krener}) showed that Kalman-Bucy filter is a minimax filter with quadratic norm of exogenous signals.  

In some applications such as in finance, the data can be very noisy and unpredictable with unstable statistical properties of the exogenous signals/noises. For such situations when there is less information about noises, worst case approaches such as $\mathcal{H}_\infty$ have been proposed (see for example \cite{grimble}, \cite{nag1} and \cite{shaked}). Nagpal and Khargonekar (\cite{nag1}) showed that the optimal fixed interval $\mathcal{H}_2$ smoother is also the optimal smoother for the worst-case $\mathcal{H}_\infty$ criteria. The proposed approach here is motivated by the well known $L_2$ worst-case optimality property of fixed interval $\mathcal{H}_2$ smoother (see for example \cite{krener} and \cite{nag1} and described in Lemma \ref{minL2} in the next section). Here we extend the $L_2$ cost criteria to two other cost functions which are suitable in presence of noisy measurements and outliers. The $\epsilon$-insensitive loss function, one of the two loss functions considered in this paper, was introduced by Vapnik and coworkers (see for example (\cite{vapnik1}) and (\cite{vapnik2})) for Support Vector Machine (SVM) algorithms in machine learning and regression. In $\epsilon$-insensitive SVM learning algorithms the cost function may be linear or quadratic and involve problem specific kernels but the common theme is that one ignores small errors which has been shown to provide more robustness and better generalizability of the algorithms. Support vector machines with such loss functions have also been applied in least squares regression and time series analysis (see for example (\cite{rojo}), \cite{muller}) and (\cite{suykens})). As illustrated in an example presented in Section 4, $\epsilon$-insensitive loss function provides good performance and smoother estimates in presence of both low frequency measurement noises (such as unknown biases) as well as high frequency noises.  

In presence of outliers and uncertainty about statistical properties of noises and disturbances, other statistical approaches such as estimation with Huber M-estimator cost function have also been proposed (see for example (\cite{Huber}, \cite{durovic} and \cite{chan}). For linear dynamical systems, estimation algorithms for handling outliers have also been proposed in \cite{aless}, \cite{andri} and \cite{garulli}. In presence of outliers, traditional estimation algorithms based on quadratic cost function can be overly influenced by outliers resulting in overall poor performance. In such cases algorithms with Huber type cost functions (\cite{Huber})  may be more appropriate as the cost function switches from quadratic to linear when errors become sufficiently large. The second cost function considered in this paper is a hybrid of Huber loss function and $\epsilon$-insensitive loss function as the one considered in (\cite{rojo}) for ARMA system identification. The proposed cost function a) ignores small measurement errors (is $\epsilon$-insensitive) and b) has linear instead of quadratic penalty for large errors making the estimates less sensitive to bias and small noises as well as outliers. 

There are instances when additional information is available about the system beyond the description of the the dynamical model. Examples of such additional information are maximum or minimum value of certain states (for example price of an asset can never be negative or physical constraints that limit movement of an object) or knowledge about possible range of changes in some states (for example a constraint on change in position over time from knowledge about maximum velocity or acceleration). One could also have some knowledge about the magnitude of disturbances and measurement noises such as an upper bound on measurement noise. Incorporating such additional information can be helpful not only in identification of outliers but can also lead to improved estimates as such information puts constraints on exogenous signals and possible trajectories of states. Thresholds for such constraints could be constants or related to given observations of the system - such as average of some states is close to the average of the measurements. One approach for estimation under such constraints is by obtaining sets constraining  possible state values (see for example \cite{milanese} and \cite{garulli}). Estimation algorithms have also been proposed when there are with equality constraints in dynamics of the system (see for example \cite{xu} and \cite{tei}). In this paper we assume the additional information about the system can be described in terms of inequality constraints that are linear with respect to states and exogenous signals. This allows for a wide variety of constraints including the examples mentioned above. Under the proposed framework, the constraints on states are not limited to ranges at a particular time but could link states at multiple periods such as constraints on average value of states over certain time interval. The constraints could also be time dependent where the constraints are different in different time intervals. Though not explicitly considered, equality constraints can also be incorporated as they can be represented as a pair of inequality constraints. The second set of results presented here provide algorithms for obtaining optimal estimates for both cost functions ($\epsilon$ insensitive quadratic as well as Huber) that also incorporate these additional constraints.

The algorithms proposed here can be applied for filtering, fixed interval smoothing as well as for prediction of future states. In all cases the estimates are obtained by solving quadratic optimization problem with linear constraints. However they cannot be recursively implemented as is the case for optimal $\mathcal{H}_2$ and $\mathcal{H}_\infty$ filtering algorithms. One other difference is that while optimal $\mathcal{H}_2$ and $\mathcal{H}_\infty$ estimates for linear systems are linear in measurements, the estimates from the proposed algorithms are not linear in measurements. However there are close similarities in the approaches - for example the proposed algorithm for $\epsilon$ insensitive quadratic loss function reduces to the optimal $\mathcal{H}_2$ and $\mathcal{H}_\infty$ smoothing algorithm when $\epsilon$ approaches zero.  

This paper is organized as follows. In the next section we describe the problems addressed in this paper and their connection to optimal $\mathcal{H}_2$ smoothing algorithm. Main results are described in Section 3 while the Appendix contains all the proofs of these results. Section 4 contains an illustrative example and the last section contains a summary of the results.

\section{Problem Formulation}

Throughout the paper, $N$ represents a positive integer which will be used to describe the number of measurements available for estimation. For vectors $v, w \in \mathbb{R}^n$, $v \geq w$ implies that all the components of the vector $v-w$ are non-negative. In particular for a real valued vector, $v \geq 0$ would imply that all elements of the vector $v$ are non-negative. For a matrix $C \in \mathbb{R}^{m \times n}$, $C'$ will indicate its transpose. For a vector $x_k \in \mathbb{R}^n$, $x_{k_j}$ denotes the $j'th$ element of $x_k$. Diagonal matrix composed of $R$ in diagonal blocks and zero everywhere else would be denoted by $diag(R)$. $I_m$ will denote identity matrix of dimension $m$.

For all the estimation problems we will assume that the underlying system is known and finite dimensional linear system of the following form: 

\begin{align}
\label{system}
    x_{k+1} & =  Ax_k+Bw_k \text{, initial condition $x_0$ is not known with $\bar{x}_0$ its best estimate} \nonumber \\
    y_k & =  Cx_k+v_k
\end{align}

where $x_k \in \mathbb{R}^n$ is the state, $y_k \in \mathbb{R}^m$ are the noisy measurements, $w_k$ and $v_k$ are unknown exogenous signals and measurement noises respectively. Given measurements $\{y_1, \hdots , y_N\}$, the prediction problem involves estimating $x_k$ for $k>N$ while for filtering the goal is to estimate $x_N$. In the problem of fixed interval smoothing, the goal is to estimate $x_k$ where $1 \leq k \leq N$. The proposed approach applies equally to linear time varying systems (when $A,B$ and $C$ depend on time index $k$ in equation (\ref{system}) but for ease of transparency, we will assume the system parameters $A \in \mathbb{R}^{n \times n}$, $B \in \mathbb{R}^{n \times l}$ and $C \in \mathbb{R}^{m \times n}$ are known constant matrices. 

For any $k \geq 0$, $\hat{x}_k$ will denote the estimate of $x_k$ based on the given measurements $\{y_1, \hdots , y_N\}$. 

\subsection{$L_2$ optimality of $\mathcal{H}_2$ and $\mathcal{H}_\infty$ smoothers}

The proposed approach is motivated by the well known $L_2$ optimality of Fixed Interval smoothing approach for both $\mathcal{H}_2$ and $\mathcal{H}_\infty$ (see for example (\cite{mayne}, \cite{krener},  \cite{weinert} and \cite{nag1}). 


\begin{lemma} \textbf{($L_2$ optimality of $\mathcal{H}_2$ and $\mathcal{H}_\infty$ smoothing)}
\label{minL2}
Let P, Q and R be positive definite weighting matrices for uncertainties in initial conditions, disturbances and measurement noise. For given measurements $\{y_1, \hdots , y_N\}$ of the system (\ref{system}), consider the following optimization problem of minimizing $L_2$ norm of disturbances and noises (a scaling parameter of $\frac{1}{2}$ is added for convenience):
\begin{equation}
    \label{minnorm}
    \argmin_{\hat{x}_0,\hat{x}_k, \hat{w}_k,\hat{v}_k} \frac{1}{2} \left[ (\hat{x}_0-\bar{x}_0)'P(\hat{x}_0-\bar{x}_0) + \sum_{k=0}^{N-1} \hat{w}_k'Q \hat{w}_k + \sum_{k=1}^{N} \hat{v}_k'R\hat{v}_k
    \right]
\end{equation}
subject to constraint that the estimated states match the given observations based on (\ref{system}), i.e
\begin{subequations}
\begin{align}
\label{constr1a}
    \hat{x}_{k+1} & =  A\hat{x}_k+B\hat{w}_k   \\
    \label{constr1b}
    y_k & =  C\hat{x}_k+\hat{v}_k
\end{align}
\end{subequations}
The solution to the optimization problem (\ref{minnorm}) subject to constraints (\ref{constr1a}) and (\ref{constr1b}) is given by the following two point boundary value problem:  
\begin{subequations}
\begin{align}
    \label{optsmoother}
    \begin{bmatrix} \hat{x}_{k+1} \\ \lambda_{k-1} \end{bmatrix} & =
    \begin{bmatrix} A & BQ^{-1}B' \\ -C'RC & A' \end{bmatrix} \begin{bmatrix} \hat{x}_{k} \\ \lambda_{k} \end{bmatrix} +
    \begin{bmatrix} 0 \\ C'R \end{bmatrix} y_k \text{ , where } \hat{x}_0=\bar{x}_0+P^{-1}A' \lambda_0 \text{ and } \lambda_N=0     \\
    \hat{w}_k & =Q^{-1}B'\lambda_k   \\
    \hat{v}_k & = y_k - C\hat{x}_k  
 \end{align}
 \end{subequations}

\end{lemma}

Though well known, for completeness a proof is provided in the Appendix. 

Mayne (\cite{mayne}) first showed the connection between $L_2$ optimality and optimal $\mathcal{H}_2$ and Bayesian smoothing estimation under the Gaussian assumption of exogenous signals. Indeed assume that $B=I_l$ and $x_0$, $w_k$ and $v_k$ are Gaussian where $w_k$ and $v_k$ are zero mean white noises with the following covariances:
 \[
   E\{x_0\}=\bar{x}_0 \ , \ E\{(x_0-\bar{x}_0)(x_0-\bar{x}_0)'\}=P^{-1}  \; , \; B=I_l , \; E\{ w_k w_j'\}=Q^{-1} \delta_{kj} \; , \; E\{v_k v_j'\}=R^{-1} \delta_{kj}
 \]
Under the above Gaussian white noise Gaussian assumptions for $w_k$ and $v_k$ it can be seen that conditional probability density is:
\begin{align}
p(x_0,.,x_N|y_1,.,y_N) & = \frac{p(x_0,\cdots,x_N) \ p(y_1, \cdots,y_N|x_0,\cdots,x_N)}{p(y_1,\cdots,y_N)}\nonumber \\
& = \frac{p(x_0) \ \prod_{k=0}^{N-1} p(x_{k+1}|x_k) \ \prod_{k=1}^{N} p(y_k|x_k)}{p(y_1,\cdots,y_N)} \nonumber \\
& = K \  \text{exp}  \left\{ -(x_0-\bar{x}_0)'P(x_0-\bar{x}_0) - \sum_{k=0}^{N-1} (x_{k+1}-Ax_k)'Q (x_{k+1}-Ax_k) - \sum_{k=1}^{N} (y_k-Cx_k)'R(y_k-Cx_k) \right\} \nonumber \\
 & = K \ \text{exp} \ \left\{ -(x_0-\bar{x}_0)'P(x_0-\bar{x}_0) - \sum_{k=0}^{N-1} w_k'Q w_k - \sum_{k=1}^{N} v_k'Rv_k \right\} \nonumber
\end{align}
where $K$ is a constant. Thus to obtain states with maximum conditional density of states, one has to minimize the cost function (\ref{minnorm}) and thus the state estimates that maximize the conditional probability density are identical to those described above in (\ref{optsmoother}).

To draw the comparison between optimal smoother above and the results presented in this paper, we will rewrite the algorithm described in optimal smoother (\ref{optsmoother}) in terms of another quadratic optimization problem. 

\begin{lemma} \textbf{(Optimal smoother as a quadratic optimization problem)}
\label{smootherconvex}
Let $M$ and $Y$ be defined as in (\ref{FGdef}) to (\ref{Mdef}) and let $\hat{\Theta} \in \mathbb{R}^{Nm}$ be the solution of the following quadratic optimization problem:

\begin{equation}
    \label{optimsmoother}
 \argmax_{\Theta} \left[ - \frac{1}{2}  \Theta' M \Theta  + \Theta' Y  \right]
\end{equation}

Let $\lambda_k$ and $\hat{x}_k$ be obtained from optimal $\hat{\Theta}$ as follows:
\begin{subequations}
\begin{align}
\label{optlambdasmoother1}
\lambda_{k-1} &  = A'\lambda_{k}+C'\hat{\theta}_k \text{ , } \lambda_N=0 \; \text{   where   }    \begin{bmatrix} \hat{\theta}_1 \\ \vdots \\ \hat{\theta}_N \end{bmatrix}=\hat{\Theta}  \text{    and   } \hat{\theta}_k \in \mathbb{R}^m \\
\label{optlambdasmoother2}
\hat{x}_{k+1} &  =  A\hat{x}_k+BQ^{-1}B'\lambda_k \text{ , } \hat{x}_0= \bar{x}_0 +P^{-1}A'\lambda_0 
\end{align}
\end{subequations}
Then $\hat{x}_k$ obtained above is the same as the optimal smoother estimate described in (\ref{optsmoother}) and $\hat{\theta}_k=R(y_k-C\hat{x}_k)$.
\end{lemma}

To see the above, note that optimal $\Theta$ in (\ref{optimsmoother}) is $\hat{\Theta}=M^{-1}Y$. Also from (\ref{Mdef}), if $Mu=z$, then one can write the relationship between $u$ and $z$ in terms of state space equations as follows where $u=\begin{bmatrix} u_1 \\ \vdots \\ u_N \end{bmatrix}$ , $z=\begin{bmatrix} z_1 \\ \vdots \\ z_N \end{bmatrix}$ and $\lambda_{k-1}=\sum_{i=k}^N A'^{(i-k)}C'u_i$ 
\begin{align}
    \lambda_{k-1} &  = A'\lambda_{k}+C'u_k \ , \ \lambda_N=0 \nonumber \\
    p_{k+1} &  =  Ap_k+BQ^{-1}B'\lambda_k \text{ , } p_0= 0 \nonumber \\
    z_k & = Cp_{k}+CA^k P^{-1} A' \lambda_0 + R^{-1}u_k \nonumber
\end{align}
With $q_k:=p_k+A^k P^{-1} A' \lambda_0$, the above set of dynamical systems can be written as
\begin{align}
    \lambda_{k-1} &  = A'\lambda_{k}+C'u_k \ , \ \lambda_N=0 \nonumber \\
    q_{k+1} &  =  Aq_k+BQ^{-1}B'\lambda_k \text{ , } q_0= P^{-1}A'\lambda_0 \nonumber \\
    z_k & = Cq_{k}+ R^{-1}u_k \nonumber
\end{align}
Thus the state space representation for $M^{-1}$ is (where $M^{-1}z=u$)
\begin{align}
    \begin{bmatrix} q_{k+1} \\ \lambda_{k-1} \end{bmatrix} & =
    \begin{bmatrix} A & BQ^{-1}B' \\ -C'RC & A' \end{bmatrix} \begin{bmatrix} q_{k} \\ \lambda_{k} \end{bmatrix} +
    \begin{bmatrix} 0 \\ C'R \end{bmatrix} z_k \text{ , where } q_0=P^{-1}A' \lambda_0 \text{ and } \lambda_N=0     \nonumber \\
    u_k & = -RCq_{k}+Rz_k  \nonumber
 \end{align}
Thus $\hat{\Theta}=M^{-1}Y$ can be written as
\begin{align}
    \begin{bmatrix} q_{k+1} \\ \lambda_{k-1} \end{bmatrix} & =
    \begin{bmatrix} A & BQ^{-1}B' \\ -C'RC & A' \end{bmatrix} \begin{bmatrix} q_{k} \\ \lambda_{k} \end{bmatrix} +
    \begin{bmatrix} 0 \\ C'R \end{bmatrix} (y_k-CA^k\bar{x}_0) \text{ , where } q_0=P^{-1}A' \lambda_0 \text{ and } \lambda_N=0     \nonumber \\
    \hat{\theta}_k & = -RC(q_{k}+A^k\bar{x}_0)+Ry_k  \nonumber
 \end{align}
 One notes from above that $\lambda_{k-1}=A'\lambda_k+C'\hat{\theta}_k$ which is the same as (\ref{optlambdasmoother1}). With $\hat{x}_k:=q_k+A^k\bar{x}_0$, the above implies that $\hat{x}_{k+1}$ also satisfies (\ref{optlambdasmoother2}) and the above can be written as
 \begin{align}
    \begin{bmatrix} \hat{x}_{k+1} \\ \lambda_{k-1} \end{bmatrix} & =
    \begin{bmatrix} A & BQ^{-1}B' \\ -C'RC & A' \end{bmatrix} \begin{bmatrix} \hat{x}_{k} \\ \lambda_{k} \end{bmatrix} +
    \begin{bmatrix} 0 \\ C'R \end{bmatrix} y_k \text{ , where } \hat{x}_0=\bar{x}_0+P^{-1}A' \lambda_0 \text{ and } \lambda_N=0     \nonumber \\
    \hat{\theta}_k & = R(y_k-C\hat{x}_k)  \nonumber
 \end{align}
The above estimate of $\hat{x}_k$ is the same as in Lemma \ref{minL2} and thus Lemmas \ref{smootherconvex} and Lemma \ref{minL2} are equivalent representations of optimal $\mathcal{H}_2$ smoother. Lemma \ref{smootherconvex} is important since all the results presented in the paper will involve optimization problems and estimates that are quite similar to (\ref{optimsmoother}), (\ref{optlambdasmoother1}) and (\ref{optlambdasmoother2}) with the main differences being additional linear terms in the cost function and/or linear constraint involving $\Theta$. 

\subsection{$\epsilon$- insensitive quadratic and Huber loss functions}

Motivated by the above $L_2$ optimality of the $\mathcal{H}_2$ smoothing algorithm, we will consider estimation problems with two objective functions in this paper. The two cost functions are $\epsilon$- insensitive quadratic loss function and $\epsilon$- insensitive Huber loss function and are illustrated in Figure 1 and described below. The loss functions are parameterized by user specified positive parameters $\epsilon$, $\kappa$ and $r$:
\begin{equation}
\label{epsiloncost}
    \epsilon \text{ insensitive quadratic loss function} \; \; \; \;  f_\epsilon(z;r,\epsilon)=
    \begin{cases}
    0 \; \text{  if  } \; |z| < \epsilon \\
   \frac{1}{2} r(|z|-\epsilon)^2 \; \text{  if  } \;  |z| \geq \epsilon 
    \end{cases}
\end{equation}

\begin{equation}
\label{hubercost}
    \epsilon \text{ insensitive Huber loss function} \; \; \; \; f_{\epsilon H}(z;r,\epsilon, \kappa)=
    \begin{cases}
    0 \; \text{  if  } \; |z| < \epsilon \\
   \frac{1}{2} r(|z|-\epsilon)^2 \; \text{  if  } \; \epsilon \leq |z| < \epsilon + \frac{\kappa}{r} \\
    \kappa(|z|-\epsilon-\frac{\kappa}{r})+\frac{\kappa^2}{2r}  \; \text{  if  } \;  |z| \geq \epsilon + \frac{\kappa}{r}
    \end{cases}
\end{equation}

In both cases the cost function is zero for sufficiently small z (for $|z| \leq \epsilon$). The difference between the above two cost functions is for larger values of $z$ (when $|z| \geq \epsilon + \frac{\kappa}{r}$) - in this case the Huber cost function is linear rather than quadratic in $z$. This feature of linear as opposed to quadratic cost function for large errors, makes the algorithm based on Huber cost function less sensitive to outliers. 

The parameters and thresholds in the above cost functions are chosen so that the cost function is continuous (has the same value at the switching points). Note that the above $\epsilon$- insensitive Huber loss function is not completely general as choices of quadratic and linear weights $r$ and $\kappa$ fix the level $\epsilon + \frac{\kappa}{r}$ where the function switches from quadratic to linear. The advantage of this functional form is that it makes the optimization problems more tractable while achieving the desired objective of linear penalty for large errors. If $\kappa$ is very large, the threshold level for switching from quadratic to linear ($\epsilon + \frac{\kappa}{r}$) would also be large and thus one would expect minimizing the above Huber cost function would lead to the same results as minimizing the $\epsilon$-insensitive quadratic loss function. This is indeed the case as described in the remark after Theorem \ref{theorem2}.

\begin{figure}
    \begin{subfigure}{0.48\textwidth}
        \includegraphics[width=\textwidth]{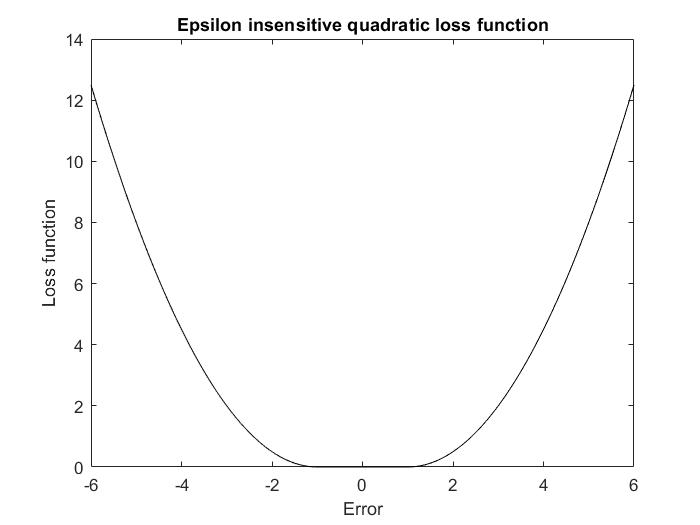}
        \caption{Quadratic $\epsilon$ insensitive loss function with $\epsilon=1$ and $r=1$}
    \end{subfigure} \hfill
    \begin{subfigure}{0.48\textwidth}
        \includegraphics[width=\textwidth]{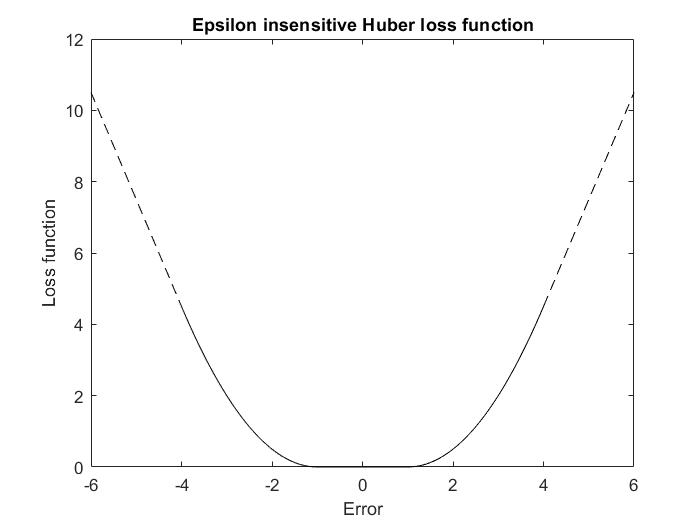}
        \caption{$\epsilon$ insensitive Huber loss function with $\epsilon=1$, $r=1$ and $\kappa=3$. The dashed line represents the region where the cost function is linear}
    \end{subfigure}
    \caption{Comparison of the two loss functions considered in this paper with $\epsilon=1$. Both loss functions are the same for $-4 \leq$Error$\leq4$ and both are zero for $-1 \leq$Error$\leq1$. For larger errors (when |Error|$>4$), Huber loss function (right plot) is linear and has smaller value than the quadratic loss function (left plot)}
 \end{figure}
 
Utilizing results from \cite{manga}, the $\epsilon$- insensitive Huber loss function can be described in terms of a convex function with linear constraints.

\begin{lemma} \textbf{($\epsilon$- insensitive Huber cost function as a convex optimization with linear constraints)}
\label{Huberopt}
The $\epsilon$- insensitive Huber cost function in (\ref{hubercost}) is given by the following optimization problem
\begin{subequations}
\begin{align}
\label{Huberopteq}
    f_{\epsilon H}(z;r,\epsilon, \kappa) = \min_{\eta_1 \ , \eta_2} \; \frac{1}{2}r \eta_1^2 + \kappa |\eta_2-\eta_1|  \\
    \text{subject to constraints } \; |z| \leq \eta_2 + \epsilon \; \text{ and } \; \ \eta_2 \geq 0
\end{align}
\end{subequations}
\end{lemma}
\emph{Proof of Lemma: } Consider the Huber cost function $f_H$ defined below without any $\epsilon$:
\[
 f_H(z;r, \kappa)=
    \begin{cases}
    \frac{1}{2} r|z|^2 \; \text{  if  } \;  |z| <  \frac{\kappa}{r} \\
    \kappa(|z|-\frac{\kappa}{r})+\frac{\kappa^2}{2r}  \; \text{  if  } \;  |z| \geq \frac{\kappa}{r}
    \end{cases}
\]
Mangasarian and Musicant \cite{manga} show that $f_H$ can be computed from the following optimization problem 
\begin{subequations}
\begin{align}
\label{mangamusic}
& \; \; \; \; \; \; \; \; \; \; f_H(z;r, \kappa)  = \min_{\eta_1} \; \frac{1}{2} r \eta_1^2 + \kappa |z-\eta_1|  \\
\label{mangarange}
& \text{Moreover optimal } \eta_1=z \text{ when } |z| <  \frac{\kappa}{r} \text{, optimal } \eta_1=\frac{\kappa}{r} \text{ when } z \geq  \frac{\kappa}{r} \text{, and optimal } \eta_1=-\frac{\kappa}{r} \text{ when } z \leq  -\frac{\kappa}{r}
\end{align}
\end{subequations}
The relationship between $\epsilon$ insensitive Huber function (\ref{hubercost}) and $f_H$ can be seen as:
\[
f_{\epsilon H}(z;r,\epsilon, \kappa)= f_H(\eta_2 ;r, \kappa) \; \; \text{where } \; \eta_2:=\text{max}(0,|z|-\epsilon)
\]
The Lemma \ref{Huberopt} follows from the above observation and (\ref{mangamusic}).

Notice that the optimization problem (\ref{Huberopteq}) can be written as the following quadratic optimization problem with linear constraints:
\begin{align}
 f_{\epsilon H}(z;r,\epsilon, \kappa) = \min_{\eta_1 \ , \eta_2 , t} \; \frac{1}{2} \ r \eta_1^2 + \kappa t  \nonumber \\
 \label{Huberconvex}
    \text{subject to constraints } \; -t \leq\eta_2-\eta_1 \leq t, \;  |z| \leq \eta_2 + \epsilon \; \text{ and } \; \ \eta_2 \geq 0 
\end{align}
The following result shows how Lemma \ref{Huberopt} can be used to convert regression problems with $\epsilon$ insensitive Huber loss function (\ref{hubercost}) into a quadratic optimization. This result may be of independent interest and is similar to many of the optimization problems we will encounter later. For the next result on regression with  $\epsilon$- insensitive Huber cost function, we will use the following definitions:
\begin{equation}
    \label{RHepskapdef}
    R_H:=\begin{bmatrix} r_1 & 0 & \cdots & 0 \\
    0 & r_2 & 0 & \cdots \\
    \cdots & \cdots & \cdots & \cdots \\
    0 & \cdots & 0 & r_m 
    \end{bmatrix} \; , \; \;
    \epsilon:=\begin{bmatrix} \epsilon_1 \\ \vdots \\ \epsilon_m \end{bmatrix}
    \; , \; \; \kappa:=\begin{bmatrix} \kappa_1 \\ \vdots \\ \kappa_m \end{bmatrix}
\end{equation}

\begin{lemma} \textbf{(Regression with $\epsilon$- insensitive Huber cost function)}
\label{Huberreg}
For given $A \in \mathcal{R}^{m \times n}$ , positive definite $Q \in \mathcal{R}^{n \times n}$ and $b \in \mathcal{R}^m$ where $m >n$ and positive parameters $r_i$, $\epsilon_i$ and $\kappa_i$, consider the regression problem of obtaining $x \in \mathcal{R}^n$ that minimizes the following $\epsilon$ insensitive Huber cost function
\begin{equation}
\label{Huberregproblem}
\min_{x \in \mathcal{R}^n} \; \left[ \frac{1}{2}x'Qx+ \sum_{i=1}^m f_{\epsilon H}((Ax-b)_i;r_i,\epsilon_i, \kappa_i) \right]
\end{equation}
where $(Ax-b)_i$ is the $i'th$ element of the vector $Ax-b$. Let $\hat{\theta}$ be the optimal $\theta$ from the following quadratic optimization problem with linear constraints:
\begin{subequations}
\begin{align}
\label{Huberregpt}
& \argmax_{\theta, \zeta} \left[-\frac{1}{2}\theta'(AQ^{-1}A'+R_H^{-1})\theta
  - \zeta'\epsilon + \theta'b  \right] \\
\label{constrHreg}
& \text{subject to constraints  } \zeta \geq \theta, \;  \zeta \geq -\theta , \; \theta \leq \kappa, \; \text{and } \; -\theta \leq \kappa
\end{align}
\end{subequations}
Then optimal $x$ for the problem (\ref{Huberregproblem}) can be obtained as
\begin{equation}
\label{xoptreg}
    x_{opt}=Q^{-1}A'\hat{\theta}
\end{equation}
\end{lemma}
For completeness, proof of the above Lemma is provided in the Appendix. It can be seen that if one removed the constraints on $\theta$ in (\ref{constrHreg}) and set $\epsilon=0$, the optimal solution $x$ above would be the same as the least squares solution.

\subsection{Estimation Problem Descriptions}

We next describe the problems addressed in this paper. In all the problems below we will assume that vector $\epsilon \in \mathbb{R}^m$ with $\epsilon>0$ is user specified constant vector.

\textbf{Problem 1 - Estimation with $\epsilon$- insensitive quadratic loss function :} Let P, Q and R be given positive definite weighting matrices for initial conditions, disturbances and measurement noise. Given measurements $\{y_1, \hdots , y_N\}$ of the system (\ref{system}), obtain the estimate from the following optimization problem :
\begin{equation}
    \label{problem1}
    \argmin_{\hat{x}_0,\hat{x}_k, \hat{w}_k, \eta_k} \frac{1}{2} \left[ (\hat{x}_0-\bar{x}_0)'P(\hat{x}_0-\bar{x}_0) + \sum_{k=0}^{N-1} \hat{w}_k'Q \hat{w}_k + \sum_{k=1}^{N} (y_k-C\hat{x}_k-\eta_k)'R(y_k-C\hat{x}_k-\eta_k)
    \right]
\end{equation}
subject to constraint that the estimated states match the given observations based on (\ref{system}) and the slack variable $\eta_k$ is less than $\epsilon$, i.e.
\begin{subequations}
\begin{align}
\label{constrdynamics}
    & \hat{x}_{k+1}  =  A\hat{x}_k+B\hat{w}_k \; , \; \forall \; k \text{ in } \{0,\cdots,N\} \\
    \label{constr2b}
   & | \eta_k |  \leq \epsilon \text{ where } \epsilon=\begin{bmatrix} \epsilon_1 \\ \vdots \\ \epsilon_m \end{bmatrix} \text{ and } \epsilon_i>0 \text{ are specified constants}
\end{align}
\end{subequations}

The key difference in the above optimization problem from the one in (\ref{minnorm}) is in the third term where $\hat{v}_k=y_k-C\hat{x}_k$ is replaced by $(y_k-C\hat{x}_k-\eta_k)$. If $|y_k-C\hat{x}_k|\leq \epsilon$, the choice of $\eta_k=y_k-C\hat{x}_k$ satisfies constraint (\ref{constr2b}) while also making the last term in the  optimization problem (\ref{problem1}) zero. This implies that there is no difference in the last term of the cost function in (\ref{problem1}) as long as $C\hat{x}_k$ stays within a tube of $\epsilon$ thickness around $y_k$. This framework of ignoring small errors in prediction estimate makes the algorithm less sensitive to measurement noises of both high frequencies as well as well as low frequencies (such as bias terms) and produces smoother estimates (Remark 1 after Theorem \ref{theorem1} also includes a comment linking the equations of the algorithm to this "smoothness" property). For a simple example of low sensitivity to measurement noises, consider a situation where $\bar{x}_0=0$, $w_k=0$, $x_k=0$, $y_k=v_k \neq 0$ and $|v_k| \leq \epsilon$. In this case while the optimal estimate from the above algorithm $\hat{x}_k$ will be $0$ (as optimal $\eta_k=y_k$), the optimal smoother described in Lemma \ref{minL2} will be nonzero (since $y_k\neq0)$. Since this approach provides smoother estimates, it is more useful in estimating states that change slowly rather than fast. An example in Section 4 illustrates improved performance of the proposed approach compared to optimal $\mathcal{H}_2$ smoother especially for slowly varying states.

The next problem considered is the estimation with $\epsilon$- insensitive Huber loss function of the form described in (\ref{hubercost}) that also provides robustness against outliers.

\textbf{Problem 2 - Estimation with $\epsilon$- insensitive Huber loss function :} Let P, Q be positive definite matrices and $\{r_1,\cdots, r_m\}$, $\{\kappa_1,\cdots, \kappa_m\}$ and $\{\epsilon_1,\cdots, \epsilon_m\}$ be positive scalar parameters associated with the Huber cost function. Given measurements $\{y_1, \hdots , y_N\}$, obtain the estimate from the following optimization problem:
\begin{equation}
    \label{problem1h}
    \argmin_{\hat{x}_0,\hat{x}_k, \hat{w}_k}  \left[ \frac{1}{2} (\hat{x}_0-\bar{x}_0)'P(\hat{x}_0-\bar{x}_0) + \frac{1}{2} \sum_{k=0}^{N-1} \hat{w}_k'Q \hat{w}_k + \sum_{k=1}^{N} \sum_{j=1}^{m} f_{Huber}\left((y_{k}-C\hat{x}_{k})_j;r_j,\epsilon_j, \kappa_j \right)
    \right]
\end{equation}
where $(y_{k}-C\hat{x}_{k})_j$ is the $j'th$ element of $(y_{k}-C\hat{x}_{k})$, $f_{Huber}$ is the $\epsilon$- insensitive Huber loss function defined in (\ref{hubercost}) and the above optimization is subject to the constraints (\ref{constrdynamics}) describing the system dynamics.

Note that the last term of the cost function (\ref{problem1h}) has $m$ terms at each time step $k$ with each corresponding to one of the $m$ measurements (recall $y_k \in \mathbb{R}^m$). This separate treatment for each measurement is in contrast to one term involving the the prediction error for each time step in (\ref{problem1}) where there is no restriction on R being diagonal (the last term with weighting matrix $R$). As shown in the comment after Theorem \ref{theorem2}, the optimization problems with quadratic and Huber cost function result in the same solution if $R$ is diagonal and $R=R_H$ and $\kappa_j$ is sufficiently large. This is due to the fact that when $\kappa_j$ is very large, the threshold error level of $\epsilon + \frac{\kappa}{r}$ when the cost function switches from quadratic to linear is also very large and thus the Huber cost function effectively becomes $\epsilon$- insensitive quadratic cost function. 

Next we will consider a class of problems where additional information is available beyond the system dynamics described in (\ref{system}) that can help in identifying noisy outliers and also enhance the accuracy of the estimate. Examples of additional information are maximum or minimum value of certain states (for example price of an asset can never be negative or physical constraints on movement of an object) or knowledge about possible range of changes in some states (for example a bound on change in position from knowledge about maximum velocity or acceleration). One could also have some knowledge about the magnitude of disturbances and measurement noises such as $|w_k|$ or $|v_k|\leq \alpha$. We will assume that the additional information is in form of linear inequality constraints where the constraints could be known constants or dependent on measurements such as average of some states is within a band around the average of measurements. Some such examples of additional information are described below:
\begin{subequations}
\begin{align}
& a  \leq \frac{1}{N} \sum_{i=1}^N L x_i \leq b \; \text{   (average value between $a$ and $b$ where these are constants or functions of measurements $y_k$)} \\
\label{measnoise}
& -Cx_i  \leq a-y_i \text{ , } Cx_i \leq a +y_i \;  \; \; \text{(measurement noise $v_i=y_i-Cx_i$ bounded by $a$)} \\
& L(x_{i+l}-x_{i})= L(A^l-I)x_i + \sum_{j=0}^{l-1} A^{l-j-1}Bw_{i+j} \leq a (y_{i+l}-y_i) +b \; \; \;  \text{   (constraint on changes over a period $l$)}
\end{align}
\end{subequations}
We will assume that the additional information about the system can be written as linear inequality constraints involving $x_k$ and $w_k$, as is the case in all the constraints described above. More specifically we will assume that the additional constraints are of the form 
\begin{equation}
\label{addconstr}
    \sum_{k=1}^NU_kx_k +  \sum_{k=0}^{N-1}V_kw_k \leq a \; \; \; \text{where $a \in \mathbb{R}^p$, $U_k \in \mathbb{R}^{p \times n}$ and $V_k \in \mathbb{R}^{p \times l}$ }
\end{equation}
In the above $a$, $U_k$ and $V_k$ are known matrices that may depend on the measurements $\{y_1, \cdots,y_N \}$ and each of the $p$ rows of the above inequality describes a different inequality constraint. For example if some state constraints are known in terms of average measurements such as $Lx_k \leq (b \frac{1}{N} \sum_1^N y_k + c ) $ for all $k$, we would have
\[
U_1=\begin{bmatrix} L \\ 0 \\ \vdots \\ 0 \end{bmatrix} \; , \; U_2=\begin{bmatrix} 0 \\ L \\ 0 \\ \vdots \end{bmatrix} \; , \;
\cdots \; , \; U_N=\begin{bmatrix} 0 \\  \vdots \\ 0 \\ L \end{bmatrix} , \;  \; V_k=0 \; \forall k \; , \; \text{and } \;  \;
a=\begin{bmatrix} b \\  \vdots \\ b \end{bmatrix} \frac{1}{N} \sum_1^N y_k + \begin{bmatrix} c \\  \vdots \\ c \end{bmatrix}
\]
Similarly if the constraints are $L(x_{k+1}-x_k)=L(A-I)x_k+LBw_k \leq a$ for all $k=1$ to $N$, we would have
\[
\begin{bmatrix} U_1 & \cdots & U_N \end{bmatrix}=\begin{bmatrix} L(A-I) & 0 & \cdots & 0 \\ 0 & L(A-I) & 0 & \cdots \\
\cdots & \cdots & \cdots & \cdots \\ 0 & \cdots & 0 & L(A-I) \end{bmatrix} , \; \; \begin{bmatrix} V_0 & \cdots & V_{N-1} \end{bmatrix} = \begin{bmatrix} LB & 0 & \cdots & 0 \\ 0 & LB & 0 & \cdots \\
\cdots & \cdots & \cdots & \cdots \\ 0 & \cdots & 0 & LB \end{bmatrix}
\]
The above is an example of constraints involving state values at different times. As another example of constraints linking states at different times, consider a constraint on average value such as $\frac{1}{N}\sum_1^N Lx_k \leq \frac{1}{N}\sum_1^N y_k+b$. In this case we would have $U_k=\frac{1}{N}L$ and $V_k=0$ for all $k$. Since $U_k$ and $V_k$ can depend on time $k$, the above formulation also allows for different constraints at different times. One also notes that though $v_k$ does not appear in the general form of constraint described in (\ref{addconstr}), constraints involving $v_k$ can also be written as above. For example $|v_k| \leq a$ can be written as in (\ref{measnoise}) since $v_k=y_k-Cx_k$. Finally, though written in form of inequalities, the proposed approach can also be applied to equality constraints by either additional equality constraint or writing equality constraint as a pair of inequalities (for example representing equality constraint $Lx=a$ as two inequality constraints $Lx \leq a$ and $-Lx \leq -a$).

For the next set of problems we consider estimation with the dual objective of minimizing $\epsilon$- insensitive loss function (both quadratic and Huber) while also incorporating such additional information about the states and exogenous signals.

\textbf{Problem 3 - Estimation with $\epsilon$- insensitive quadratic loss function and constraints : } Let P, Q and R be given positive definite weighting matrices for uncertainties in initial conditions, disturbances and measurement noise. Given measurements $\{y_1, \hdots , y_N\}$ of the system (\ref{system}), obtain the estimate from the following optimization problem :
\begin{equation}
    \label{problem2}
    \argmin_{\hat{x}_0,\hat{x}_k, \hat{w}_k} \frac{1}{2} \left[ (\hat{x}_0-\bar{x}_0)'P(\hat{x}_0-\bar{x}_0) + \sum_{k=0}^{N-1} \hat{w}_k'Q \hat{w}_k + \sum_{k=1}^{N} (y_k-C\hat{x}_k-\eta_k)'R(y_k-C\hat{x}_k-\eta_k)
    \right]
\end{equation}
subject to constraints in Problem 1 ((\ref{constrdynamics}) and (\ref{constr2b})) and the additional constraints about the system:
\begin{align}
    \label{constr3c}
    \sum_{k=1}^NU_k\hat{x}_k & +  \sum_{k=0}^{N-1}V_k\hat{w}_k \leq a \; \; \; \text{where $a$, $U_k$ and $V_k$ are known given measurements  } \{y_1, \cdots,y_N \}
\end{align}

The next problem considers estimation with the same constraints but with Huber loss function.

\textbf{Problem 4 - Estimation with $\epsilon$- insensitive Huber loss function and constraints : } Let P, Q and R be given positive definite matrices, $\{r_1,\cdots, r_m\}$, $\{\kappa_1,\cdots, \kappa_m\}$ and $\{\epsilon_1,\cdots, \epsilon_m\}$ be positive scalar parameters associated with the Huber cost function. Given measurements $\{y_1, \hdots , y_N\}$ of the system (\ref{system}), obtain the estimate from the following optimization problem :
\begin{equation}
    \label{problem4}
    \argmin_{\hat{x}_0,\hat{x}_k, \hat{w}_k}  \left[ \frac{1}{2} (\hat{x}_0-\bar{x}_0)'P(\hat{x}_0-\bar{x}_0) + \frac{1}{2} \sum_{k=0}^{N-1} \hat{w}_k'Q \hat{w}_k + \sum_{k=1}^{N}  \sum_{j=1}^{m} f_{Huber}\left((y_{k}-C\hat{x}_{k})_j;r_j,\epsilon_j, \kappa_j \right)
    \right]
\end{equation}
where $(y_{k}-C\hat{x}_{k})_j$ is the $j'th$ element of $(y_{k}-C\hat{x}_{k})$ and $f_{Huber}$ is the $\epsilon$- insensitive Huber loss function defined in (\ref{hubercost}). The above optimization is subject to constraint related to given dynamics (\ref{constrdynamics}) and additional constraints about the system (\ref{constr3c}).

In all the problems described above, the goal is to estimate $\{ \hat{x}_0, \cdots, \hat{x}_N \}$ given the measurements $\{y_1, \cdots, y_N \}$. In the problem of prediction we are interested in estimating future states for which we do not yet have any measurements - that is estimate $\hat{x}_{N+j}$ for some $j \geq 1$ while the available measurements are only $\{y_1, \cdots, y_N \}$. Problems 5 and 6 described below are a natural extension of the the proposed framework for estimating future states with the same two cost functions considered above.
 
 \textbf{Problem 5 - Prediction with $\epsilon$- insensitive quadratic loss function and constraints :} Let P, Q and R be given positive definite weighting matrices for uncertainties in initial conditions, disturbances and measurement noise. Given measurements $\{y_1, \hdots , y_N\}$ of the system (\ref{system}), determine the estimate $\hat{x}_{N+j}$ for a given $j \geq 1$ from the following optimization problem:
\begin{equation}
    \label{problempred}
    \argmin_{\hat{x}_0,\hat{x}_k, \hat{w}_k} \frac{1}{2} \left[ (\hat{x}_0-\bar{x}_0)'P(\hat{x}_0-\bar{x}_0) + \sum_{k=0}^{N+j-1} \hat{w}_k'Q \hat{w}_k + \sum_{k=1}^{N} (y_k-C\hat{x}_k-\eta_k)'R(y_k-C\hat{x}_k-\eta_k)
    \right]
\end{equation}
subject to constraints (\ref{constrdynamics}), (\ref{constr2b}) and the following: 
\begin{align}
    \label{constrpred}
    \sum_{k=1}^{N+j}U_k \hat{x}_k & +  \sum_{k=0}^{N+j-1}V_k \hat{w}_k \leq a \; \; \; \text{where $a$, $U_k$ and $V_k$ are known given measurements  } \{y_1, \cdots,y_N \}
\end{align}

The next problem considers the problem of prediction with constraints and $\epsilon$- insensitive Huber loss function.

 \textbf{Problem 6 - Prediction with $\epsilon$- insensitive Huber loss function and constraints :} Let P, Q and R be given positive definite matrices, $\{r_1,\cdots, r_m\}$, $\{\kappa_1,\cdots, \kappa_m\}$ and $\{\epsilon_1,\cdots, \epsilon_m\}$ be positive scalar parameters associated with the Huber cost function. Given measurements $\{y_1, \hdots , y_N\}$ of the system (\ref{system}), determine the estimate $\hat{x}_{N+j}$ for a given $j \geq 1$ from the following optimization problem:
\begin{equation}
    \label{problempredH}
    \argmin_{\hat{x}_0,\hat{x}_k, \hat{w}_k} \left[ \frac{1}{2} (\hat{x}_0-\bar{x}_0)'P(\hat{x}_0-\bar{x}_0) + \frac{1}{2} \sum_{k=0}^{N+j-1} \hat{w}_k'Q \hat{w}_k + \sum_{k=1}^{N}  \sum_{j=1}^{m} f_{Huber}\left((y_{k}-C\hat{x}_{k})_j;r_j,\epsilon_j, \kappa_j \right)
    \right]
\end{equation}
where $(y_{k}-C\hat{x}_{k})_j$ is the $j'th$ element of $(y_{k}-C\hat{x}_{k})$ and $f_{Huber}$ is the $\epsilon$- insensitive Huber loss function defined in (\ref{hubercost}). The above optimization is subject to constraints (\ref{constrdynamics}) and (\ref{constrpred}). 

There are two main differences in Problems $5$ and $6$ compared to Problems $1$ to $4$ - a) the objective functions (\ref{problempred}) and (\ref{problempredH}) include contribution of exogenous signal $\hat{w}_k$ for all $k \leq N+j-1$ (instead of just $k \leq N-1$) and, b) the constraints (\ref{constrpred}) incorporates all constraints involving $\hat{w}_k$ and $\hat{x}_k$ that influence all the states up to $\hat{x}_{N+j}$ (in other words last two problems also include constraints beyond the last measurement period $N$).

\section{Main Results}

We next describe solutions to the six problems described above. In the results below please notice the similarity to the optimal fixed interval smoothing algorithm as described in Lemma \ref{smootherconvex}. The following definitions will be used in describing the results:


The following definitions will be used in describing the results:
\begin{subequations}
\begin{align}
\label{FGdef}
  &   F  :=\begin{bmatrix} CB & 0 & \cdots & 0 \\
 CAB & CB & 0 & \cdots  \\
 \cdots & \cdots & \cdots & \cdots \\
 CA^{N-2}B & \cdots & CB & 0 \\
 CA^{N-1}B & \cdots & CAB & CB \end{bmatrix} \;  ,  \;
Y  :=\begin{bmatrix} y_1-CA\bar{x}_0 \\ \vdots \\ y_N-CA^N\bar{x}_0 \end{bmatrix} \; , \;
Q_{inv}:= \begin{bmatrix} Q^{-1} & 0 & \cdots & 0 \\
 0 & Q^{-1} & 0 & \cdots  \\
 \cdots & \cdots & \cdots & \cdots \\
 0 & \cdots & 0 & Q^{-1} \end{bmatrix}
  \\
 \label{XYdef}
 &   R_{inv}:= \begin{bmatrix} R^{-1} & 0 & \cdots & 0 \\
 0 & R^{-1} & 0 & \cdots  \\
 \cdots & \cdots & \cdots & \cdots \\
 0 & \cdots & 0 & R^{-1} \end{bmatrix} \; 
  \; , \; R_{Hinv}:= \begin{bmatrix} R_H^{-1} & 0 & \cdots & 0 \\
 0 & R_H^{-1} & 0 & \cdots  \\
 \cdots & \cdots & \cdots & \cdots \\
 0 & \cdots & 0 & R_H^{-1} \end{bmatrix} \; \text{where $R_H$ is defined in (\ref{RHepskapdef}) }
        \\
      \label{epskappa}
     & \varepsilon:=\begin{bmatrix} I_m \\ \vdots \\ I_m \end{bmatrix} \epsilon \; , \; 
     K:=\begin{bmatrix} I_m \\ \vdots \\ I_m \end{bmatrix} \kappa \; , \; \text{where $\epsilon$ and $\kappa$ are defined in (\ref{RHepskapdef}) and }  \; \varepsilon  , \ K \in \mathbb{R}^{mN} \\
     \label{Mdef}
     &   M := FQ_{inv}F' + R_{inv} + \begin{bmatrix} CA \\ \vdots \\ CA^{N} \end{bmatrix} P^{-1}\begin{bmatrix} A'C' & \hdots & A'^{N}C' \end{bmatrix} \\
      \label{MHdef}
     &   M_H := FQ_{inv}F' + R_{Hinv} + \begin{bmatrix} CA \\ \vdots \\ CA^{N} \end{bmatrix} P^{-1}\begin{bmatrix} A'C' & \hdots & A'^{N}C' \end{bmatrix}
 \end{align}
 \end{subequations}

The result below shows that the optimal solution to Problem 1 described by (\ref{problem1}) subject to constraints (\ref{constrdynamics}) and (\ref{constr2b}) can be obtained by solving a standard quadratic optimization with linear constraints.

\begin{theorem}
\label{theorem1}
(Problem $1$) : Let $\hat{\Theta} , \hat{\zeta} \in \mathbb{R}^{Nm}$ be the solution of the following quadratic optimization problem with linear constraints:

\begin{subequations}
\begin{align}
\label{opteps}
& \argmax_{\Theta, \zeta} \left[ - \frac{1}{2}  \Theta' M \Theta - \varepsilon \zeta  + \Theta' Y  \right] \\
\label{constreps}
& \text{subject to constraints  } \zeta \geq \Theta \text{  and  } \zeta \geq -\Theta  
\end{align}
\end{subequations}

Let $\lambda_k$ be obtained from optimal $\hat{\Theta}$ as follows:

\begin{equation}
\label{optlambdaep}
\lambda_{k-1}  = A'\lambda_{k}+C'\hat{\theta}_k \text{ , } \lambda_N=0 \; \text{   where   }    \begin{bmatrix} \hat{\theta}_1 \\ \vdots \\ \hat{\theta}_N \end{bmatrix}=\hat{\Theta}  \text{    and   } \hat{\theta}_k \in \mathbb{R}^m
\end{equation}
With $\lambda_k$ obtained as above, the optimal values for $\hat{x}_k$, $\hat{w}_k$ that minimize (\ref{problem1}) are obtained as follow:
\begin{subequations}
\begin{align}
\label{optxpr1}
& \hat{x}_{k+1}  =  A\hat{x}_k+BQ^{-1}B'\lambda_k \text{ , } \hat{x}_0= \bar{x}_0 +P^{-1}A'\lambda_0 \\
& \hat{w}_k=Q^{-1}B'\lambda_k 
\end{align}
 \end{subequations}
 \end{theorem}
 
Proof of the above Theorem is provided in the Appendix.
 
 \emph{Remark 1}: Comparing the above to Lemma \ref{smootherconvex}, one notes that a) the expressions for $\lambda_i$ and $\hat{x}_i$ above are the same as (\ref{optlambdasmoother1}) and (\ref{optlambdasmoother2}), and b) if $\epsilon=0$, the optimization problem (\ref{opteps}) becomes the same as the optimization problem (\ref{optimsmoother}). The difference in the optimization problem (\ref{opteps}) compared to (\ref{optlambdasmoother1}) is the additional term $\varepsilon' \zeta =\varepsilon' |\Theta|$ (since optimal $\zeta=|\Theta|$). This additional $l_1$ penalty results in smaller optimal $|\hat{\theta}_k|$ for optimization problem (\ref{opteps}) compared to the same in (\ref{optimsmoother}). Since $\lambda_i$ and $\hat{x}_i$ are outputs of the same linear system in both cases ((\ref{optlambdaep}) and (\ref{optxpr1})) with $\hat{\theta}_k$ as the input, the "smaller" $|\hat{\theta}_k|$ produces smaller changes in $\lambda_k$ and $\hat{x}_k$ in Theorem \ref{theorem1} compared to Lemma \ref{smootherconvex} and thus results in smoother estimates of states. The greater smoothness of estimates, which can be qualitatively understood as a consequence of being tolerant for certain level of error, will also be illustrated in the example described below.

The next result provides solution to Problem 2 which is the estimation problem with $\epsilon$ insensitive Huber loss function.

\begin{theorem}
\label{theorem2}
(Problem $2$): Let $\hat{\Theta} , \hat{\zeta} \in \mathbb{R}^{Nm}$ be the solution of the following quadratic optimization problem with linear constraints:

\begin{subequations}
\begin{align}
\label{optepsH}
& \argmax_{\Theta, \zeta} \left[ - \frac{1}{2}  \Theta' M_H \Theta   - \varepsilon \zeta  + \Theta' Y  \right] \\
\label{constrepsH}
& \text{subject to constraints  } \zeta \geq \Theta, \; \zeta \geq -\Theta  , \; \Theta \leq K \text{  and  }  -\Theta  \leq K
\end{align}
\end{subequations}

where $\varepsilon$ and $K$ are as defined in (\ref{epskappa}). Let $\lambda_k$ be obtained from optimal $\hat{\Theta}$ as follows:

\begin{equation}
\label{optlambdaepH}
\lambda_{k-1}  = A'\lambda_{k}+C'\hat{\theta}_k \text{ , } \lambda_N=0 \; \text{   where   }    \begin{bmatrix} \hat{\theta}_1 \\ \vdots \\ \hat{\theta}_N \end{bmatrix}=\hat{\Theta}  \text{    and   } \hat{\theta}_k \in \mathbb{R}^m
\end{equation}
With $\lambda_k$ obtained as above, the optimal values for $\hat{x}_k$, $\hat{w}_k$ that minimize (\ref{problem1h}) are obtained as follow:
\begin{subequations}
\begin{align}
\label{optxpr1H}
& \hat{x}_{k+1}  =  A\hat{x}_k+BQ^{-1}B'\lambda_k \text{ , } \hat{x}_0= \bar{x}_0 +P^{-1}A'\lambda_0 \\
& \hat{w}_k=Q^{-1}B'\lambda_k 
\end{align}
 \end{subequations}
 \end{theorem}
 
 \emph{Remark 2}: Notice again the similarity between (\ref{optepsH}) and (\ref{optsmoother}). The primary difference between Theorem \ref{theorem1} and \ref{theorem2} is the additional constraint of $|\Theta| \leq K$ (which is the same as $|\theta_{k_j}|\leq \kappa_j$ for all $k$ and $j$). One key observation from comparing Theorems \ref{theorem1} and \ref{theorem2} is that the optimal $\hat{\Theta}$ and thus the optimal estimates $\hat{x}_k$ are the same for both if $R=R_H=diag(r_i)$ and $K$ (or equivalently $\kappa$) is very large. To see this note that $M=M_H$ when $R=R_H$ and when $K$ is very large, the constraint $|\Theta| \leq K$ is not active and thus optimal $\hat{\Theta}$ in Theorems \ref{theorem1} and \ref{theorem2} are the same. For intuition behind this observation note that $\epsilon$-insensitive Huber cost function effectively becomes $\epsilon$-insensitive quadratic cost function when $\kappa$ is very large as the threshold $\epsilon_j+\frac{\kappa_j}{r_j}$ for switch from quadratic to linear is not reached. In the subsequent results as well we will notice that the estimates from both $\epsilon$-insensitive quadratic and Huber cost functions are the same if $R=R_H$ and $K$ (or equivalently $\kappa$) is very large.

For the next result, define
\begin{equation}
    \label{VGHdef}
    V:=\begin{bmatrix} V_0 & \cdots & V_{N-1} \end{bmatrix}, \; 
    G  :=\begin{bmatrix} B' & B'A' & \cdots & B'A'^{N-1} \\ 0 & B' & \cdots & B'A'^{N-2} \\ \cdots & \cdots & \cdots & \cdots \\
 0 & \cdots & 0 & B' \end{bmatrix} \begin{bmatrix} U_1' \\ \vdots \\ U_N' \end{bmatrix},
 H:= \begin{bmatrix} CA \\ \vdots \\ CA^N \\   -\sum_{i=1}^{N} U_iA^i \end{bmatrix} 
\end{equation}
\begin{align}
\label{Tdef}
    T:=\begin{bmatrix} F \\ -(G'+V) \end{bmatrix} Q_{inv} \begin{bmatrix} F' & -(G+V') \end{bmatrix} + \begin{bmatrix}
    R_{inv} & 0 \\ 0 & 0 \end{bmatrix}+HP^{-1}H' \\
    \label{THdef}
    T_H:=\begin{bmatrix} F \\ -(G'+V) \end{bmatrix} Q_{inv} \begin{bmatrix} F' & -(G+V') \end{bmatrix} +\begin{bmatrix}
    R_{Hinv} & 0 \\ 0 & 0 \end{bmatrix}+ HP^{-1}H'
\end{align}

The following result for Problem 3 describes the algorithm to obtain the optimal estimate for the $\epsilon$ insensitive quadratic loss function (\ref{problem2}) subject to additional knowledge about the system in terms of constraints defined in (\ref{constr3c}).
 
\begin{theorem}
\label{theorem3}
(Problem 3): Let $\hat{\Theta} , \hat{\zeta} \in \mathbb{R}^{Nm}$ and $\hat{\xi} \in \mathbb{R}^{p}$ be the solution of the following quadratic optimization problem with linear constraints:

\begin{subequations}
\begin{align}
\label{opteps2}
\argmax_{\Theta, \zeta, \xi} & \left[ - \frac{1}{2} \begin{bmatrix} \Theta & \xi \end{bmatrix}' T \begin{bmatrix} \Theta \\ \xi \end{bmatrix} 
-  \varepsilon' \zeta + \Theta' Y - \xi'(a-\sum_{i=1}^NU_iA^i\bar{x}_0) \right] \\
\label{constreps2}
& \text{subject to constraints  } \zeta \geq \Theta \text{  ,  } \zeta \geq -\Theta  \; \; \text{and } \xi \geq0
\end{align}
\end{subequations}

Let $\lambda_k$ be obtained from optimal $\hat{\Theta}$ and $\hat{\xi}$ as follows

\begin{equation}
\label{optlambdaep2}
\lambda_{k-1}  = A'\lambda_{k}+C'\hat{\theta}_k -U_k' \hat{\xi} \; \text{ , } \lambda_N=0 \; \text{   where   }    \begin{bmatrix} \hat{\theta}_1 \\ \vdots \\ \hat{\theta}_N \end{bmatrix}=\hat{\Theta}  \text{    and   } \hat{\theta}_k \in \mathbb{R}^m
\end{equation}
With $\lambda_k$ obtained as above, the optimal values for $\hat{x}_k$, $\hat{w}_k$ that minimize (\ref{problem2}) subject to constraints (\ref{constrdynamics}), (\ref{constr2b}) and (\ref{constr3c}) are obtained as follow:
\begin{subequations}
\begin{align}
\label{optxpr2}
& \hat{x}_{k+1}  =  A\hat{x}_k+BQ^{-1}B'\lambda_k -BQ^{-1}V_k' \hat{\xi} \; , \; \; \hat{x}_0= \bar{x}_0 +P^{-1}A'\lambda_0 \\
& \hat{w}_k=Q^{-1}B'\lambda_k - Q^{-1}V_k' \hat{\xi} \nonumber
\end{align}
 \end{subequations}
 \end{theorem}
 
 The next result for Problem 4 describes the algorithm to obtain the optimal estimate for the $\epsilon$ insensitive Huber loss function subject to additional knowledge about the system in terms of constraints defined in (\ref{constr3c}).
 
 \begin{theorem}
\label{theorem4}
(Problem 4): Let $\hat{\Theta} , \hat{\zeta} \in \mathbb{R}^{Nm}$ and $\hat{\xi} \in \mathbb{R}^{p}$ be the solution of the following quadratic optimization problem with linear constraints:

\begin{subequations}
\begin{align}
\label{opteps2H}
\argmax_{\Theta, \zeta, \xi} & \left[ - \frac{1}{2} \begin{bmatrix} \Theta & \xi \end{bmatrix}' T_H \begin{bmatrix} \Theta \\ \xi \end{bmatrix} 
-  \varepsilon' \zeta + \Theta' Y - \xi'(a-\sum_{i=1}^NU_iA^i\bar{x}_0) \right] \\
\label{constreps2H}
& \text{subject to constraints  } \zeta \geq \Theta \text{  ,  } \zeta \geq -\Theta . \; \Theta \leq K, \;  -\Theta \leq K, \text{and } \xi \geq0
\end{align}
\end{subequations}

where $\varepsilon$ and $K$ are as defined in (\ref{epskappa}). Let $\lambda_k$ be obtained from optimal $\hat{\Theta}$ and $\hat{\xi}$ as follows

\begin{equation}
\label{optlambdaep2H}
\lambda_{k-1}  = A'\lambda_{k}+C'\hat{\theta}_k -U_k' \hat{\xi} \; \text{ , } \lambda_N=0 \; \text{   where   }    \begin{bmatrix} \hat{\theta}_1 \\ \vdots \\ \hat{\theta}_N \end{bmatrix}=\hat{\Theta}  \text{    and   } \hat{\theta}_k \in \mathbb{R}^m
\end{equation}
With $\lambda_k$ obtained as above, the optimal values for $\hat{x}_k$, $\hat{w}_k$ that minimize (\ref{problem4}) subject to constraints (\ref{constrdynamics}) and (\ref{constr3c}) are obtained as follow:
\begin{subequations}
\begin{align}
\label{optxpr2H}
& \hat{x}_{k+1}  =  A\hat{x}_k+BQ^{-1}B'\lambda_k -BQ^{-1}V_k' \hat{\xi} \; , \; \; \hat{x}_0= \bar{x}_0 +P^{-1}A'\lambda_0 \\
& \hat{w}_k=Q^{-1}B'\lambda_k - Q^{-1}V_k' \hat{\xi} \nonumber
\end{align}
 \end{subequations}
 \end{theorem}
 
 Note that the difference in optimization problems of Theorems \ref{theorem3} and \ref{theorem4} is the additional constraint of $|\Theta| \leq K$ in Theorem \ref{theorem4} (similar to the difference in optimization problems of Theorems \ref{theorem1} and \ref{theorem2}). 
 
 For describing the results of the prediction problems (Problems 5 and 6) where the goal is to estimate $\hat{x}_{N+j}$ where $j \geq 1$ given available measurements only up to $N$, we will use the following notation:
 
 \begin{subequations}
\begin{align}
\label{FGdefa}
 & \bar{F} :=\begin{bmatrix} F & 0_{Nm \times jl} \end{bmatrix} \;  ,  \; \text{where } 0_{Nm \times jl} \in \mathbb{R}^{Nm \times jl} \; \text{where each element of the matrix is $0$} \\
 \label{FGdefb}
 & \bar{G}  :=\begin{bmatrix} B' & B'A' & \cdots & B'A'^{N+j-1} \\ 0 & B' & \cdots & B'A'^{N+j-2} \\ \cdots & \cdots & \cdots & \cdots \\
 0 & \cdots & 0 & B' \end{bmatrix} \begin{bmatrix} U_1' \\ \vdots \\ U_{N+j}' \end{bmatrix} \; , \;
\bar{V}:=\begin{bmatrix} V_0 & \cdots & V_{N+j-1} \end{bmatrix}, \; \;
\bar{H}:= \begin{bmatrix} CA \\ \vdots \\ CA^N \\   -\sum_{i=1}^{N+j} U_iA^i \end{bmatrix} \\
& \bar{Q}_{inv}:= \begin{bmatrix} Q^{-1} & 0 & \cdots & 0 \\
 0 & Q^{-1} & 0 & \cdots  \\
 \cdots & \cdots & \cdots & \cdots \\
 0 & \cdots & 0 & Q^{-1} \end{bmatrix} \;, \; \text{(a diagonal matrix with $Q^{-1}$ in diagonals)}  
\end{align}
\end{subequations}

\begin{align}
\label{Tdefa}
   \bar{T}:=\begin{bmatrix} \bar{F} \\ -(\bar{G}'+\bar{V}) \end{bmatrix} \bar{Q}_{inv} \begin{bmatrix} \bar{F}' & -(\bar{G}+\bar{V}') \end{bmatrix} + \begin{bmatrix}
    R_{inv} & 0 \\ 0 & 0 \end{bmatrix}+\bar{H}P^{-1}\bar{H}' \\
    \label{TdefaH}
   \bar{T}_H:=\begin{bmatrix} \bar{F} \\ -(\bar{G}'+\bar{V}) \end{bmatrix} \bar{Q}_{inv} \begin{bmatrix} \bar{F}' & -(\bar{G}+\bar{V}') \end{bmatrix} +  \begin{bmatrix}
    R_{Hinv} & 0 \\ 0 & 0 \end{bmatrix} + \bar{H}P^{-1}\bar{H}'
\end{align}
 
 The following result for Problem 5 describes the algorithm to obtain the optimal prediction for the $\epsilon$ insensitive quadratic loss function (\ref{problempred}) subject to additional knowledge about the system (\ref{constrpred}).
 
\begin{theorem}
\label{theorem5}
(Problem 5): Let $\hat{\Theta} , \hat{\zeta} \in \mathbb{R}^{Nm}$ and $\hat{\xi} \in \mathbb{R}^{p}$ be the solution of the following quadratic optimization problem with linear constraints:

\begin{subequations}
\begin{align}
\label{opteps3}
\argmax_{\Theta, \zeta, \xi} & \left[ - \frac{1}{2} \begin{bmatrix} \Theta & \xi \end{bmatrix}' \bar{T} \begin{bmatrix} \Theta \\ \xi \end{bmatrix} 
-  \varepsilon' \zeta + \Theta' Y - \xi'(a-\sum_{i=1}^{N+j}U_iA^i\bar{x}_0) \right] \\
\label{constreps3}
& \text{subject to constraints  } \zeta \geq \Theta \text{  ,  } \zeta \geq -\Theta  \; \; \text{and } \xi \geq0
\end{align}
\end{subequations}

Let $\lambda_k$ be obtained from optimal $\hat{\Theta}$ and $\hat{\xi}$ as follows

\begin{subequations}
\begin{align}
\label{optlambdaep3a}
\lambda_{k-1}  & = A'\lambda_{k} -U_k' \hat{\xi} \; ,  \; \text{   with   } \lambda_{N+j}=0 , \; \;\text{for   } N+1 \leq k \leq N+j  \\
\label{optlambdaep3b}
\lambda_{k-1}  & = A'\lambda_{k}+C'\hat{\theta}_k -U_k' \hat{\xi} \; , \;\text{for   } 1 \leq k \leq N \; \;  \text{   where   }    \begin{bmatrix} \hat{\theta}_1 \\ \vdots \\ \hat{\theta}_N \end{bmatrix}=\hat{\Theta}  \text{    and   } \hat{\theta}_k \in \mathbb{R}^m 
\end{align}
\end{subequations}
With $\lambda_k$ obtained as above, the optimal values for $\hat{x}_k$, $\hat{w}_k$ that minimize (\ref{problempred}) subject to constraints (\ref{constrdynamics}), (\ref{constr2b}) and (\ref{constrpred}) are obtained as follow:

\begin{subequations}
\begin{align}
\label{optxpr3}
& \hat{x}_{k+1}  =  A\hat{x}_k+BQ^{-1}B'\lambda_k -BQ^{-1}V_k' \hat{\xi} \; , \; \; \hat{x}_0= \bar{x}_0 +P^{-1}A'\lambda_0 \\
& \hat{w}_k=Q^{-1}B'\lambda_k - Q^{-1}V_k' \hat{\xi} \nonumber
\end{align}
 \end{subequations}
 \end{theorem}
 
 The next result describes the solution to the prediction problem with $\epsilon$ insensitive Huber cost function (Problem 6).
 
 \begin{theorem}
\label{theorem6}
(Problem 6): Let $\hat{\Theta} , \hat{\zeta} \in \mathbb{R}^{Nm}$ and $\hat{\xi} \in \mathbb{R}^{p}$ be the solution of the following quadratic optimization problem with linear constraints:

\begin{subequations}
\begin{align}
\label{opteps3H}
\argmax_{\Theta, \zeta, \xi} & \left[ - \frac{1}{2} \begin{bmatrix} \Theta & \xi \end{bmatrix}' \bar{T}_H \begin{bmatrix} \Theta \\ \xi \end{bmatrix} 
 -  \varepsilon' \zeta + \Theta' Y - \xi'(a-\sum_{i=1}^{N+j}U_iA^i\bar{x}_0) \right] \\
\label{constreps3H}
& \text{subject to constraints  } \zeta \geq \Theta \text{  ,  } \zeta \geq -\Theta, \; \Theta \leq K, \; -\Theta \leq K \; \text{and } \xi \geq0
\end{align}
\end{subequations}

where $\varepsilon$ and $K$ are as defined in (\ref{epskappa}). Let $\lambda_k$ be obtained from optimal $\hat{\Theta}$ and $\hat{\xi}$ as follows

\begin{subequations}
\begin{align}
\label{optlambdaep3aH}
\lambda_{k-1}  & = A'\lambda_{k} -U_k' \hat{\xi} \; ,  \; \text{   with   } \lambda_{N+j}=0 , \; \;\text{for   } N+1 \leq k \leq N+j  \\
\label{optlambdaep3bH}
\lambda_{k-1}  & = A'\lambda_{k}+C'\hat{\theta}_k -U_k' \hat{\xi} \; , \;\text{for   } 1 \leq k \leq N \; \;  \text{   where   }    \begin{bmatrix} \hat{\theta}_1 \\ \vdots \\ \hat{\theta}_N \end{bmatrix}=\hat{\Theta}  \text{    and   } \hat{\theta}_k \in \mathbb{R}^m 
\end{align}
\end{subequations}
With $\lambda_k$ obtained as above, the optimal values for $\hat{x}_k$, $\hat{w}_k$ that minimize (\ref{problempredH}) subject to constraints (\ref{constrdynamics}) and (\ref{constrpred}) are obtained as follow:

\begin{subequations}
\begin{align}
\label{optxpr3H}
& \hat{x}_{k+1}  =  A\hat{x}_k+BQ^{-1}B'\lambda_k -BQ^{-1}V_k' \hat{\xi} \; , \; \; \hat{x}_0= \bar{x}_0 +P^{-1}A'\lambda_0 \\
& \hat{w}_k=Q^{-1}B'\lambda_k - Q^{-1}V_k' \hat{\xi} \nonumber
\end{align}
 \end{subequations}
 \end{theorem}
 
 For all the results described above the dimension of the variables and thus the computational requirements continue to grow as the size of the data $N$ grows. Unlike recursive Kalman-Bucy and $\mathcal{H}_\infty$ filtering algorithms, to obtain the estimate $\hat{x}_{N+1}$ with the addition of new data $y_{N+1}$, one has to recompute all the states $\{ \hat{x}_0, \cdots, \hat{x}_{N+1} \}$. In real time applications of filtering and prediction, this might be computationally prohibitive. In such cases a practical way to use the proposed approach would be to use the most recent $N$ observations where $N$ is chosen as large as computationally feasible.
 
 \section{Illustrative Example}
 
For illustrative example, we will consider discrete version of mass spring damper system:
 \[
 m \ddot{x} \ + \ b \dot{x}  \ + \ k x  \ = \ w
 \]
 where $w$ is unknown random exogenous signal and the available measurements of the position can be very noisy. One reason for considering this system is that the two states have different rates of variation (frequencies) - while the position $x$ changes smoothly, the velocity $\dot{x}$ can change quickly if the exogenous signal $w$ has power spectrum in high frequencies. Below it will be observed that the $\epsilon$ insensitive feature of ignoring small measurement errors leads to smoother estimates where the performance of the proposed algorithms is better in estimation of position (lower frequency states) compared to velocity (higher frequency states). With states $x_1=x$ and $x_2=\dot{x}$ with time  step of $\Delta t$, the above can be written as discrete model of the form (\ref{system}) where : 
  \[
 A=\begin{bmatrix} 1 & \Delta t \\ -\frac{k}{m} \Delta t & -\frac{b}{m} \Delta t \end{bmatrix} \; , \; B=\begin{bmatrix} 0 \\ 1 \end{bmatrix} \; , \;  C=\begin{bmatrix} 1 & 0 \end{bmatrix}
 \]
For the example the assumed parameters and time step are $m=3$, $b=2$, $k=2$ and $\Delta t=0.5$. Initial conditions and the weights for initial condition, exogenous signal, $\epsilon$ and $\kappa$ parameters for the two cost functions are:
 \[
x_0=\begin{bmatrix} -1 \\ 1 \end{bmatrix} \; , \; \bar{x}_0=\begin{bmatrix} 0 \\ 0 \end{bmatrix} \; , \; Q=1 \; , \; R=1 \; , \;  P=\begin{bmatrix} 1 & 0 \\ 0 & 1 \end{bmatrix} \; , \; \epsilon= \{2.5 \ , \ 5\} \; , \; \kappa=4
 \]
Two values for $\epsilon$ are considered to see its sensitivity on performance. For the simulation exercises, it is assumed that exogenous signals and measurement noises are of the form 
 \begin{align}
 w_k & =  5 \ r_{1k} \; ,  \; \text{where $r_{1k} \sim \mathcal{N}(0,1) $} \nonumber \\
 v_k & = \begin{cases} 5 \ r_{2k} \ + \ 6 \;, \text{with probability $0.8$}  \; \text{where $r_{2k} \sim \mathcal{N}(0,1)$ } \nonumber \\
 20 \ r_{2k} \ + \ 6 \;, \text{with probability $0.2$}  \; \text{where $r_{2k} \sim \mathcal{N}(0,1)$ } \nonumber \\
 \end{cases} 
 \end{align}
where $\mathcal{N}(0,1)$ denotes normal random variable with mean of zero and standard deviation of one. Note that the measurement noise $v_k$ has bias of $6$ and to represent outliers, measurement noise becomes large with probability $0.2$ when the amplitude multiplying the normal random variable jumps from $5$ to $20$. 

Simulations were performed under these assumptions for thirty time steps ($N=30$). Table \ref{example1} shows the comparison of Root Mean Square Error (RMSE) and Mean Absolute Error (MAE) absolute errors for the proposed estimation algorithms with $\mathcal{H}_2$ smoothing algorithm described in Lemma \ref{minL2}. Here the error measures RMSE and MAE  are defined as follows:
\begin{align}
    \text{RMSE error for for first state $x_1$} & = \text{average over simulations of } \; \sqrt{\frac{1}{N+1}  \sum_{k=0}^N (x_{1_k}-\hat{x}_{1_k})^2} \nonumber \\
    \text{MAE error for for first state $x_1$} & = \text{average over simulations of } \; \frac{1}{N+1}  \sum_{k=0}^N |x_{1_k}-\hat{x}_{1_k}| \nonumber
\end{align}
Error metrics for the second state are defined similarly. The Table \ref{example1} describes comparison of RMSE and MAE (averaged over a number of simulations) for optimal $\mathcal{H}_2$ smoother, $\epsilon$-insensitive quadratic as well as Huber cost function approaches (algorithms described in Theorems \ref{theorem1} and \ref{theorem2}). The proposed algorithms based on $\epsilon$-insensitive quadratic and Huber cost function clearly perform better than optimal $\mathcal{H}_2$ smoothing algorithm in estimation of position (smaller error for $x_1$). In estimation of velocity which has higher frequency changes compared to position, the $\epsilon$ insensitive quadratic cost function performance is slightly worse than that of optimal $\mathcal{H}_2$ smoother. This is due to the fact that tolerance of small measurement errors results in smoother estimates and thus higher frequency changes are not captured as well by $\epsilon$ insensitive approaches. Because of presence of outliers, $\epsilon$ insensitive Huber cost function estimates are the best in estimating both position and velocity. 

\begin{table}
\begin{center}
\begin{tabular}{|l |c|c|c| } 
 \hline
 & Optimal $\mathcal{H}_2$ smoother & $\epsilon$-insensitive quadratic & $\epsilon$-insensitive Huber \\ 
  \hline
   & & $\epsilon=2.5 \; \; | \; \; \epsilon=5$ & $\epsilon=2.5 \; \; | \; \; \epsilon=5$ \\
   \hline
RMSE $x_1$ (position) & 6.39 & 6.27 \; \; | \; \; 6.01  & 5.55 \; \; | \; \; 5.37   \\ 
\hline
MAE $x_1$ (position) & 5.5 & 5.31 \; \; | \; \; 4.98 &  4.74 \; \; | \; \; 4.36 \\ 
\hline
\hline
RMSE $x_2$ (velocity) & 5 & 5.05 \; \; | \; \; 5.17 &  4.67 \; \; | \; \; 4.83 \\ 
 \hline
MAE $x_2$  (velocity) & 3.98 & 4.02 \; \; | \; \; 4.16 &  3.74 \; \; | \; \; 3.87 \\
\hline
\end{tabular}
\end{center}
\caption{Comparison of RMSE and MAE for position and velocity for the mass spring damper system. Estimation errors for position ($x_1$) are lower with $\epsilon$ insensitive quadratic and Huber cost functions compared to optimal $\mathcal{H}_2$ smoother. In estimation of velocity ($x_2$ which has faster changes), $\epsilon$ insensitive quadratic estimator performance is slightly worse than that of optimal $\mathcal{H}_2$ smoother.}
\label{example1}
\end{table}

Next we extend this example to estimation in presence of constraints while all the other assumptions about exogenous signals, noises and the dynamical system are the same as above. For the constraint, we will assume that it is known that the absolute value of velocity cannot exceed $4$, i.e.
 \[
| \begin{bmatrix} 0 & 1 \end{bmatrix} x_k | \leq 4  \; \; \forall k \; \; \text{(assumed known constraint on velocity)}
\]
Table \ref{example2} shows the comparison of RMSE and MAE for the proposed $\epsilon$-insensitive estimation algorithms based on Theorems \ref{theorem3} and \ref{theorem4} with optimal $\mathcal{H}_2$ smoothing algorithm. Here the improved performance of the $\epsilon$-insensitive algorithms is not only due to lower sensitivity to measurement noises but also due to the ability to incorporate constraints in the estimates. Figure 2 shows the comparison of actual value as well as estimates using different algorithms for one sample path. As discussed before, one observes that the estimates from both $\epsilon$-insensitive quadratic and Huber cost function are smoother than those obtained using optimal $\mathcal{H}_2$ smoother.

\begin{table}
\begin{center}
\begin{tabular}{|l |c|c|c| } 
 \hline
 & Optimal $\mathcal{H}_2$ smoother & $\epsilon$-insensitive quadratic & $\epsilon$-insensitive Huber \\ 
  \hline
   & & $\epsilon=2.5 \; \; | \; \; \epsilon=5$ & $\epsilon=2.5 \; \; | \; \; \epsilon=5$ \\
   \hline
RMSE $x_1$ (position) & 6.40 & 5.68 \; \; | \; \; 5.51  & 5.09 \; \; | \; \; 4.91   \\ 
\hline
MAE $x_1$ (position) & 5.60 & 5.03 \; \; | \; \; 4.71 &  4.54 \; \; | \; \; 4.21 \\ 
\hline
\hline
RMSE $x_2$ (velocity) & 4.30 & 3.22 \; \; | \; \; 3.36 &  2.98 \; \; | \; \; 3.12 \\ 
 \hline
MAE $x_2$  (velocity) & 3.45 & 2.47 \; \; | \; \; 2.61 &  2.34 \; \; | \; \; 2.48 \\
\hline
\end{tabular}
\end{center}
\caption{Comparison of RMSE and MAE for position and velocity for the mass spring damper system with the constraint $|x_2(k)| \leq 4$ for all $k$.}
\label{example2}
\end{table}

\begin{figure}
 \label{examplefig2}
    \begin{subfigure}{0.48\textwidth}
        \includegraphics[width=\textwidth]{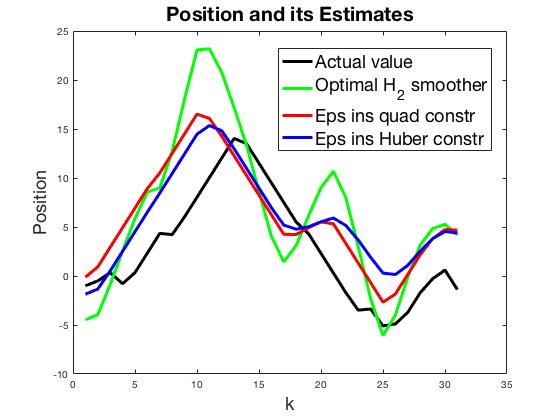}
    \end{subfigure} \hfill
    \begin{subfigure}{0.48\textwidth}
        \includegraphics[width=\textwidth]{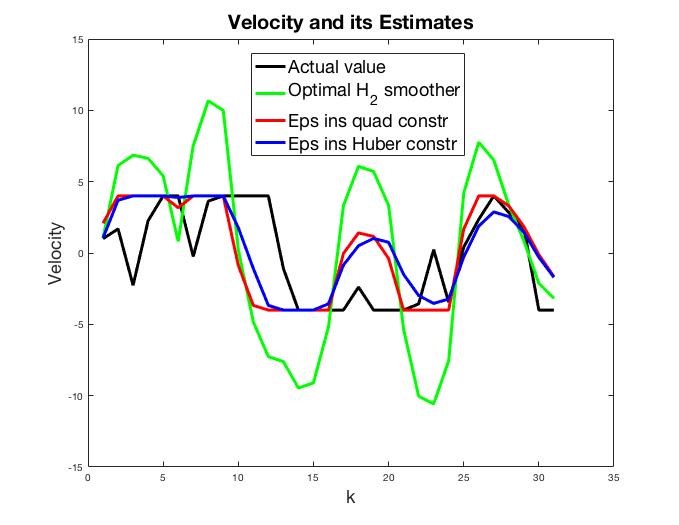}
    \end{subfigure}
    \caption{Comparison of actual value, estimates using optimal $\mathcal{H}_2$ smoother, optimal $\epsilon$ insensitive quadratic and Huber loss function estimators for one sample path under the constraint that $|x_2(k)| \leq 4$ for all $k$. }
 \end{figure}

\section{Summary}
This paper presents optimal estimation algorithms for linear dynamical systems with $\epsilon$-insensitive loss functions. Such an optimization criteria is often used in Support Vector Machines and provides greater robustness and lower sensitivity to measurement noises as small errors are ignored. One of the cost functions is $\epsilon$-insensitive Huber cost function for which the penalty function switches from quadratic to linear for large errors which makes the estimates less sensitive to outliers. We also present algorithms to estimate states with the same objective functions while also incorporating additional constraints about the system. Results are also provided for the prediction problem where the goal is to estimate future states of the system. Though not recursive, the algorithms are are easily implemented as they involve solving quadratic optimization problems with linear constraints. An example illustrates the improved performance of the proposed algorithms compared to Kalman-Bucy or optimal $\mathcal{H}_2$ smoothing algorithm.

\appendix

\section{Appendix}
This first part of the Appendix provides proof of Lemmas described in the Problem Formulation section (Lemmas (\ref{minL2}) and (\ref{Huberreg})). The proofs of estimation problems with $\epsilon$ insensitive quadratic cost function (Theorems \ref{theorem1}, \ref{theorem3} and \ref{theorem5}) are provided next as they are similar to each other. Proofs of algorithms with $\epsilon$ insensitive Huber cost function are provided at the end of this section.

\subsection{Proof of Lemma \ref{minL2}}
Substituting for $\hat{v}_k=y_k-C\hat{x}_k$, the Lagrangian for the optimization problem (\ref{minnorm}) subject to constraints (\ref{constr1a}) and (\ref{constr1b}) is
\[
L= \frac{1}{2} \left[ (\hat{x}_0-\bar{x}_0)'P(\hat{x}_0-\bar{x}_0) + \sum_{k=0}^{N-1} \hat{w}_k'Q \hat{w}_k + \sum_{k=1}^{N} (y_k-C\hat{x}_k)'R(y_k-C\hat{x}_k)  \right] + \sum_{k=0}^{N} \lambda_k' \left(\hat{x}_{k+1} -  A\hat{x}_k-B\hat{w}_k  \right)
\]
where $\lambda_k \in \mathbb{R}^n$ for $k=0$ to $N$ are the Lagrange multipliers. The necessary conditions for minimum which are also sufficient given convex cost function and linear constraints are 
\begin{subequations}
\begin{align}
    \label{necconds1}
    \frac{\delta L}{\delta \hat{x}_k} & =0 \text{ for 1$\leq$ k $\leq$ N} \Rightarrow C'RC\hat{x}_k-C'Ry_k+\lambda_{k-1}-A'\lambda_k =0 
    \Rightarrow \lambda_{k-1} = A'\lambda_k - C'RC \hat{x}_k + C'Ry_k  \\
    \frac{\delta L}{\delta \hat{x}_0} & =0 \Rightarrow P\hat{x}_0 - P\bar{x}_0 - A' \lambda_0=0  \nonumber \Rightarrow \hat{x}_0 =\bar{x}_0 +P^{-1}A'\lambda_0  \\
    \frac{\delta L}{\delta \hat{x}_{N+1}} & =0 \Rightarrow  \lambda_N=0  \\
    \label{necconds1a}
    \frac{\delta L}{\delta \hat{w}_k} & =0  \Rightarrow  Q\hat{w}_k - B'\lambda_k=0 \Rightarrow \hat{w}_k=Q^{-1}B'\lambda_k   \\
    \label{necconds1b}
    \frac{\delta L}{\delta \lambda_k} & =0  \Rightarrow \hat{x}_{k+1} =  A\hat{x}_k+B\hat{w}_k 
\end{align}
\end{subequations}
From equations (\ref{necconds1a}) and (\ref{necconds1b}), one obtains $\hat{x}_{k+1} =  A\hat{x}_k+BQ^{-1}B'\lambda_k$. Optimal smoother described in (\ref{optsmoother}) follows from the above optimality conditions.

\subsection{Proof of Lemma \ref{Huberreg}}

From Lemma (\ref{Huberopt}), the optimal $x$ for problem (\ref{Huberregproblem}) can be obtained from the following optimization problem
\begin{align}
    \min_{x,\eta_1, \eta_2, t} \left[\frac{1}{2} x'Qx + \frac{1}{2}  \sum_{i=1}^m r_i \eta_{1_i}^2  
    +  \sum_{i=1}^m \kappa_i t_i \right] \nonumber \\
    \text{subject to constraints } \; |Ax-b| \leq \varepsilon + \eta_2 \;, \;  \eta_2 \geq 0 \; \; \text{and } \; |\eta_{1} - \eta_{2}| \leq t 
\end{align}
The above optimization problem can be written in terms of its Lagrangian as
\begin{align}
 \max_{\gamma_1,\gamma_2, \beta_1, \beta_2, \mu} \; \min_{x, \eta_1, \eta_2, t} \;&  \Big[ \frac{1}{2} x'Qx + \frac{1}{2} \eta_1'R_H \eta_1 
    + \kappa' t  - \gamma_1'(\epsilon + \eta_2 - Ax+b) - \gamma_2'(\epsilon + \eta_2 + Ax-b)  \nonumber \\
  &  - \beta_1'(t-\eta_1 +\eta_2) - \beta_2'(t+\eta_1 -\eta_2)
  - \mu'\eta_2 \Big]  \nonumber
\end{align}
\[
    \text{         with  } \gamma_1 \geq 0 , \; \gamma_2 \geq 0, \;  \beta_1 \geq 0, \; \beta_2 \geq  0 \; \mu \geq  0 \;\;
    \; \text{(due to inequality constraints)} 
 \]
 With $L$ defined as the Lagrangian expression above,  Karush-Kuhn-Tucker (KKT) conditions for optimality are 
\begin{subequations}
\begin{align}
     \frac{\delta L}{\delta x} & =0  \Rightarrow 
    x=Q^{-1}A'(\gamma_2-\gamma_1)  \\
    \label{eta1reg}
     \frac{\delta L}{\delta \eta_{1}} & =0  \Rightarrow 
    \eta_{1}=R_H^{-1}(\beta_{2}-\beta_{1})  \\
    \label{gambetamu}
     \frac{\delta L}{\delta \eta_{2}} & =0  \Rightarrow \gamma_{1}+ \gamma_{2}+\beta_{1}-\beta_{2} +\mu  =0    \\
     \frac{\delta L}{\delta t} & =0 \Rightarrow 
    \beta_{1} + \beta_{2}= \kappa   \\
     \gamma_{1}, & \; \gamma_{2}, \; \beta_{1}, \; \beta_{2}, \; \mu \;\geq 0   \\
    \gamma_1' & (\epsilon + \eta_2 - Ax+b)=0  \; , \;\gamma_2'(\epsilon + \eta_2 + Ax-b)=0 \; , \; \beta_1'(t-\eta_1 +\eta_2)=0 \; , \; \beta_2'(t+\eta_1 -\eta_2) \; , \; \mu'\eta_2 =0 
\end{align}
\end{subequations}

Utilizing the above optimality conditions, we next show two facts :
\begin{align}
& \gamma_{1_i}+ \gamma_{2_i} = \beta_{2_i}-\beta_{1_i}=|\gamma_{1_i} - \gamma_{2_i}| \; \; \text{for all }  i \nonumber \\
& -\kappa_i \leq \gamma_{1_i} - \gamma_{2_i} \leq \kappa_i \; \; \text{for all }  i \nonumber
\end{align}

If $\eta_{2_i} > 0$ then $\mu_i=0$ and thus from (\ref{gambetamu}) one observes that $\gamma_{1_i}+ \gamma_{2_i} = \beta_{2_i}-\beta_{1_i}$. If $\eta_{2_i}=0$ then $\eta_{1_i}=0$ (since from (\ref{mangarange}) $\eta_{1_i}=\eta_{2_i}$ if $|\eta_{2_i}| \leq \frac{\kappa_i}{r_i}$). From the optimal expression for $\eta_1$ in (\ref{eta1reg}), we conclude that $\beta_{2_i}-\beta_{1_i}=0$ when $\eta_{1_i}=0$ since $R_H$ is diagonal. Thus in this case also $\gamma_{1_i}+ \gamma_{2_i}=\beta_{2_i}-\beta_{1_i}$ since $0 \leq \gamma_{1_i}+ \gamma_{2_i} \leq \beta_{2_i}-\beta_{1_i}=0$ (from (\ref{gambetamu}) since $\mu_i \geq 0$). Note that from KKT optimality condition if $\gamma_{1_i} \neq 0 $ then $\gamma_{2_i}=0$ and vice versa since $\epsilon_i >0$ and $\eta_{2_i} \geq 0$ (because  $\gamma_{1_i}(\epsilon + \eta_2 - Ax+b)_i=0$ and $\gamma_{2_i}(\epsilon + \eta_2 + Ax-b)_i=0$). Thus $\gamma_{1_i}+ \gamma_{2_i}=|\gamma_{2_i}- \gamma_{1_i}|$ for all $i$. To see the second claim above let us consider two separate cases of $t_i=0$ and $t_i>0$. In the first case $t_i=0$ implies $\eta_{1_i}=\eta_{2_i}$. From Lemma 2.2 equation (\ref{mangarange}) we know that $\eta_{1_i}=\eta_{2_i}$ when $|\eta_{1_i}| \leq \frac{\kappa_i}{r_i}$. From the optimal expression of $\eta_{1_i}$ above, this implies $|\beta_{2_i}-\beta_{1_i}| \leq \kappa_i$. When $t_i>0$ at least one of $\beta_{1_i}$ or $\beta_{2_i}$ has to be zero and thus $|\beta_{2_i}-\beta_{1_i}|=\beta_{1_i} + \beta_{2_i}=\kappa_i$. Thus $|\beta_{2_i}-\beta_{1_i}| \leq \kappa_i$ in all cases. Since $\eta_{2_i}>0$ (and thus $\mu_i=0$) when $t_i >0$, one notes that $\beta_{2_i}-\beta_{1_i}=\gamma_{1_i}+ \gamma_{2_i} \geq 0$, $\beta_{2_i}-\beta_{1_i}=|\beta_{2_i}-\beta_{1_i}| \leq \kappa_i$. From this and  the first claim above one notes that $-\kappa_i \leq \gamma_{1_i} - \gamma_{2_i} \leq \kappa_i$. Utilizing these two facts, the Lagrangian for the optimization problem is
\begin{align}
L & =-\frac{1}{2}(\gamma_2-\gamma_1)'AQ^{-1}A'(\gamma_2-\gamma_1)
-\frac{1}{2}(\beta_{2}-\beta_{1})'R_H^{-1}(\beta_2-\beta_1)  - (\gamma_1+\gamma_2)'\varepsilon + (\gamma_2-\gamma_1)'b \nonumber \\
& = -\frac{1}{2}(\gamma_2-\gamma_1)'AQ^{-1}A'(\gamma_2-\gamma_1)
-\frac{1}{2}(\gamma_{2}-\gamma_{1})'R_H^{-1}(\gamma_2-\gamma_1)  - (\gamma_1+\gamma_2)'\varepsilon + (\gamma_2-\gamma_1)'b \nonumber \\
& = -\frac{1}{2}\theta'(AQ^{-1}A'+R_H^{-1})\theta
  - |\theta|'\varepsilon + \theta'b \; \; \text{ where } \theta:=\gamma_2-\gamma_1 \nonumber
\end{align}
The second equality above holds due to the fact that $\frac{1}{r_i}(\beta_{2_i}-\beta_{1_i})^2=\frac{1}{r_i}(\gamma_{2_i}-\gamma_{1_i})^2$. As shown above $-\kappa_i \leq \theta_i \leq \kappa_i$ and with $\zeta:=|\theta|$, one obtains the results described in Lemma \ref{Huberreg}.

\subsection{Proof of Theorem \ref{theorem1}}
The optimization problem (\ref{problem1}) subject to constraints (\ref{constrdynamics}) to (\ref{constr2b}) can be described as follows in terms of its Lagrangian where $\lambda_k \in \mathbb{R}^n$, $\gamma_k$ and $\beta_k \in \mathbb{R}^m$ are the Lagrange multipliers :

\begin{align}
 & \max_{\lambda_k,\gamma_k, \beta_k} \text{   } \min_{\hat{x}_0,\hat{x}_k, \hat{w}_k, \eta_k} \;  L \;  \; \; \; \; \; \text{         with  } \gamma_k \geq 0 \text{ and } \beta_k \geq 0   \nonumber \\
 \label{lagrangpr1}
\text{   where  } \; & L  =  \frac{1}{2} \left[ (\hat{x}_0-\bar{x}_0)'P(\hat{x}_0-\bar{x}_0) + \sum_{k=0}^{N-1} \hat{w}_k'Q \hat{w}_k + \sum_{k=1}^{N} (y_k-C\hat{x}_k-\eta_k)'R(y_k-C\hat{x}_k-\eta_k)  \right]  \nonumber \\
& + \sum_{k=0}^{N} \lambda_k' \left(\hat{x}_{k+1} -  A\hat{x}_k-B\hat{w}_k  \right)  - \sum_{k=1}^{N} \gamma_k'(\epsilon-\eta_k) - \sum_{k=1}^{N} \beta_k'(\epsilon+\eta_k) 
\end{align}

In the above, $\gamma_k \geq 0$ and $\beta_k \geq 0$ because of the constraint (\ref{constr2b}). The necessary conditions for optimality are
\begin{subequations}
\begin{align}
\label{lambdaeq1}
    \frac{\delta L}{\delta \hat{x}_k} & =0 \text{ for 1$\leq$ k $\leq$ N} \Rightarrow C'RC\hat{x}_k-C'R(y_k-\eta_k)+\lambda_{k-1}-A'\lambda_k =0 
    \Rightarrow \lambda_{k-1} = A'\lambda_k - C'RC \hat{x}_k + C'R(y_k-\eta_k)  \\
    \label{initcondn1}
    \frac{\delta L}{\delta \hat{x}_0} & =0 \Rightarrow P\hat{x}_0 - P\bar{x}_0 - A' \lambda_0=0  
    \Rightarrow \hat{x}_0 =\bar{x}_0 +P^{-1}A'\lambda_0  \\
    \label{lambdabound}
    \frac{\delta L}{\delta \hat{x}_{N+1}} & =0 \Rightarrow  \lambda_N=0  \\
    \label{weq1}
    \frac{\delta L}{\delta \hat{w}_k} & =0  \Rightarrow  Q\hat{w}_k - B'\lambda_k=0 \Rightarrow \hat{w}_k=Q^{-1}B'\lambda_k  \\
    \label{etaeq1}
    \frac{\delta L}{\delta \eta_k} & =0  \Rightarrow  R\eta_k - R(y_k-C\hat{x}_k)+(\gamma_k-\beta_k)=0 
    \Rightarrow \eta_k = (y_k-C\hat{x}_k) -R^{-1}(\gamma_k-\beta_k)  \\
    \label{xeq}
    \gamma_k & \geq 0, \; \gamma_k'(\epsilon-\eta_k)=0 ,\; \beta_k \geq 0, \; \beta_k'(\epsilon+\eta_k)=0 \; \text{  and  } \frac{\delta L}{\delta \lambda_k} =0 \Rightarrow
    \hat{x}_{k+1} =  A\hat{x}_k+B\hat{w}_k
\end{align}
\end{subequations}
Define
\begin{equation}
    \label{thetadef}
    \theta_k:=\gamma_k-\beta_k \text{ ,  } \Theta=\begin{bmatrix} \theta_1 \\ \vdots \\ \theta_N \end{bmatrix}
\end{equation}
Let $\gamma_k'=[\gamma_{k_1} \cdots \gamma_{k_m}]$, $\beta_k'=[\beta_{k_1} \cdots \beta_{k_m}]$ and $(\epsilon-\eta_k)_i$ be the $i'th$ element of $(\epsilon-\eta_k)$. From KKT optimality conditions as in the proof of Lemma \ref{Huberreg}, we know that $\gamma_{k_i}(\epsilon-\eta_k)_i=0$ and $\beta_{k_i}(\epsilon+\eta_k)_i=0$  $\forall$ $i$ and $k$. Thus $\gamma_{k_i} \neq  0$ only if $(\epsilon-\eta_k)_i=0$ and $\beta_{k_i} \neq  0$ only if $(\epsilon+\eta_k)_i=0$. Since $\epsilon >0$, it follows that if $\gamma_{k_i} \neq  0$ then $\beta_{k_i} =  0$ and if $\beta_{k_i} \neq  0$ then $\gamma_{k_i} =0$. Since 
$\gamma_{k_i}, \beta_{k_i} \geq  0$ and at most only one of them can be non-zero, one observes
\begin{equation}
    \label{thetamax}
    \gamma_{k_i} + \beta_{k_i}= |\gamma_{k_i} - \beta_{k_i}| \text{ for all $i=1 \cdots m$  and $k=1 \cdots N$ } \Rightarrow \gamma_k+\beta_k=|\theta_k| \text{  where  } |\theta_k| = \begin{bmatrix} |\theta_{k_1}| \\ \vdots \\ |\theta_{k_m}| \end{bmatrix}
\end{equation}
From (\ref{lambdaeq1}), (\ref{lambdabound}) and (\ref{etaeq1}) and the fact that $\theta_k=\gamma_k-\beta_k$, one obtains the expression for optimal $\lambda_k$ described in (\ref{optlambdaep}). Similarly one obtains optimal expression for obtaining $\hat{x}_k$ described in (\ref{optxpr1}) from (\ref{weq1}) and (\ref{xeq}). One also notes that

\begin{subequations}
\begin{align}
\label{lambdaeq}
& \lambda_{k-1}  = A'\lambda_{k}+C'\theta_k \text{ , } \lambda_N=0 \Rightarrow \begin{bmatrix} B'\lambda_0 \\ \vdots \\ B'\lambda_{N-1} \end{bmatrix}= F' \Theta \text{   where  F is defined in (\ref{FGdef})  } \\
& \hat{x}_{k+1}  =  A\hat{x}_k+BQ^{-1}B'\lambda_k \text{ , } \hat{x}_0= \bar{x}_0 +P^{-1}A'\lambda_0 \Rightarrow  
\begin{bmatrix} C\hat{x}_1 \\ \vdots \\ C\hat{x}_N \end{bmatrix} = F Q_{inv} \begin{bmatrix} B'\lambda_0 \\ \vdots \\ B'\lambda_{N-1} \end{bmatrix} + \begin{bmatrix} CA \\ \vdots \\ CA^N
\end{bmatrix}  (\bar{x}_0 +P^{-1}A'\lambda_0) \nonumber \\
\end{align}
\end{subequations}
From the above and noting from that $\lambda_0=[C' \; A'C' \; \cdots \; A'^{N-1}C']\Theta$ from (\ref{lambdaeq}), one observes that 
\begin{align}
\label{Chatx}
&  \Theta'\begin{bmatrix} C\hat{x}_1 \\ \vdots \\ C\hat{x}_N \end{bmatrix} = \Theta'FQ_{inv} F' \Theta
 + \Theta'\begin{bmatrix} CA \\ \vdots \\ CA^{N} \end{bmatrix} P^{-1}A'\begin{bmatrix} C' & \hdots & A'^{N-1}C' \end{bmatrix} \Theta +\Theta'\begin{bmatrix} CA \\ \vdots \\ CA^N \end{bmatrix}  \bar{x}_0  \\
 \label{wnormlambda}
 & \sum_{k=0}^{N-1} \lambda_k'B'Q^{-1}B \lambda_k =\Theta' FQ_{inv}F' \Theta
 \end{align}

Incorporating the above together with (\ref{initcondn1}) to (\ref{etaeq1}), the definition of $\theta_k=\gamma_k-\beta_k$ and the fact that $\gamma_k+\beta_k=|\theta_k|$, the Lagrangian in equation (\ref{lagrangpr1}) becomes

\begin{subequations}
\begin{align}
L & = \frac{1}{2} \left[ \lambda_0' A P^{-1} A' \lambda_0 + \sum_{k=0}^{N-1} \lambda_k'B'Q^{-1}B \lambda_k 
+ \sum_{k=1}^{N} (\gamma_k-\beta_k)'R^{-1}(\gamma_k-\beta_k) \right] -  \epsilon' \sum_{k=1}^{N} (\gamma_k+\beta_k) \nonumber \\
& - \sum_{k=1}^{N} (\gamma_k-\beta_k)'R^{-1}(\gamma_k-\beta_k) + \sum_{k=1}^{N} (\gamma_k-\beta_k) (y_k-C\hat{x}_k) \nonumber \\
& = \frac{1}{2} \left[  \lambda_0' A P^{-1} A' \lambda_0 + \Theta'FQ_{inv}F'\Theta + \Theta'R_{inv}\Theta \right] 
- \epsilon' \sum_{k=1}^{N} |\theta_k| - \Theta'R_{inv}\Theta  - \Theta' \begin{bmatrix} C\hat{x}_1 \\ \vdots \\ C\hat{x}_N \end{bmatrix} + \Theta' \begin{bmatrix} y_1 \\ \vdots \\ y_N \end{bmatrix} \nonumber \\
& = - \frac{1}{2}  \Theta' M \Theta - \epsilon' \sum_{k=1}^{N} |\theta_k|  + \Theta' Y \; \text{ where $M$ and $Y$ are as defined in (\ref{Mdef}) and (\ref{FGdef}) } \nonumber
\end{align}
\end{subequations}
In obtaining the last expression above, we utilized the fact that $\lambda_0=[ C' \; A'C' \; \cdots \; A'^{N-1}C'] \Theta$ and the equation (\ref{Chatx}). Since $\epsilon > 0$ and $|\theta_k| \geq0$, one notes that for any $\Theta$,
\[
 - \frac{1}{2}  \Theta' M \Theta - \epsilon' \sum_{k=1}^{N} |\theta_k|  + \Theta' Y = \max_\zeta
 \left[ - \frac{1}{2}  \Theta' M \Theta - \varepsilon' \zeta  + \Theta' Y \right] \; \; \text{subject to constraints } \zeta \geq \Theta, \zeta \geq -\Theta
 \]
where $\varepsilon$ is defined in (\ref{epskappa}). From above one notes that the dual of the optimization problem  (\ref{problem1}) subject to constraints (\ref{constrdynamics}) and (\ref{constr2b}) is as described by (\ref{opteps}) subject to constraint (\ref{constreps}) (since optimal $\zeta=|\Theta|$).

\subsection{Proof of Theorem \ref{theorem2}} 


Utilizing Lemma \ref{Huberopt} we can write the Huber loss expression in terms of the following optimization problem with variables $\eta_{1_k}, \eta_{2_k}\in \mathbb{R}^m$ and $s_k=|\eta_{1_k}- \eta_{2_k}|$ for each $k \in \{1,\cdots,N\}$:
\begin{subequations}
\begin{align}
\sum_{k=1}^{N} \sum_{j=1}^{m} f\left((y_{k_j}-C\hat{x}_{k_j});r_j,\epsilon_j, \kappa_j \right)
& =\min_{\eta_{1_k}  ,  \eta_{2_k}} \; \sum_{k=1}^{N} \sum_{j=1}^{m} \left[ \frac{1}{2} r_j \eta_{1_{k_j}}^2 + \kappa_j |\eta_{2_{k_j}}-\eta_{1_{k_j}}| \right]  \\
\label{Hubercostproof}
& =\min_{\eta_{1_k}    \eta_{2_k}   s_k} \; \sum_{k=1}^{N} \sum_{j=1}^{m} \left[ \frac{1}{2} r_j \eta_{1_{k_j}}^2 + \kappa_j s_{k_j} \right]
\end{align}
\end{subequations}
subject to the constraints
\begin{align}
    (y_{k}-C\hat{x}_{k})_j & \leq \epsilon_j + \eta_{2_{k_j}} \; \text{ for all } k \in \{1,\cdots,N\} \; \text{ and }  j \in \{1,\cdots,m\} \nonumber \\
    -(y_{k}-C\hat{x}_{k})_j & \leq \epsilon_j + \eta_{2_{k_j}}  \; \text{ for all } k \in \{1,\cdots,N\} \; \text{ and }  j \in \{1,\cdots,m\} \nonumber \\
    \eta_{2_k} & \geq 0 \; \text{ for all } k \in \{1,\cdots,N\} \nonumber \\
    s_{k_j} & \geq \eta_{2_{k_j}}-\eta_{1_{k_j}} \; \text{ for all } k \in \{1,\cdots,N\} \; \text{ and }  j \in \{1,\cdots,m\} \nonumber \\
    s_{k_j} & \geq \eta_{1_{k_j}}-\eta_{2_{k_j}} \; \text{ for all } k \in \{1,\cdots,N\} \; \text{ and }  j \in \{1,\cdots,m\} \nonumber
\end{align}
Incorporating the above, the Lagrangian for the optimization problem (\ref{problem1h}) can be written as
\begin{align}
 & \max_{\lambda_k,\gamma_{1_k}, \gamma_{2_k}, \beta_{1_k}, \beta_{2_k}, \varphi_k} \text{   } \; \; \min_{\hat{x}_0,\hat{x}_k, \hat{w}_k, \eta_{k_1} , \eta_{k_2}, s_k} \;  L \;  \; \; \; \; \; \text{ subject to constraints  } \ \gamma_{1_k}, \gamma_{2_k}. \beta_{1_k}, \beta_{2_k}, \varphi_k  \geq 0 \nonumber \\
 \label{lagrangpr1H}
\text{   where  } \; & L  =   \left[ \frac{1}{2} (\hat{x}_0-\bar{x}_0)'P(\hat{x}_0-\bar{x}_0) + \frac{1}{2} \sum_{k=0}^{N-1} \hat{w}_k'Q \hat{w}_k + \frac{1}{2} \sum_{j=1}^{m} \sum_{k=1}^N r_j\eta_{1_{k_j}}^2 +
\sum_{j=1}^{m} \sum_{k=1}^N \kappa_j s_{k_j}  \right] + \sum_{k=0}^{N} \lambda_k' \left(\hat{x}_{k+1} -  A\hat{x}_k-B\hat{w}_k  \right) \nonumber \\
&  - \sum_{k=1}^{N} \left[ \gamma_{1_k}'(\epsilon+\eta_{2_k}-y_k+C\hat{x}_k) +  \gamma_{2_k}'(\epsilon+\eta_{2_k}+y_k-C\hat{x}_k) + 
\beta_{1_k}'(s_k+\eta_{1_k}-\eta_{2_k}) +
\beta_{2_k}'(s_k-\eta_{1_k}+\eta_{2_k}) +
\varphi_k' \eta_{2_k} \right]
\end{align}
The constraints om $\gamma_{1_k}, \gamma_{2_k}. \beta_{1_k}, \beta_{2_k}, \varphi_k \geq 0$ are due to the corresponding inequality constraints. The optimality conditions for the Lagrangian above are 
\begin{subequations}
\begin{align}
\label{necHeta1}
& \frac{\delta L}{\delta \eta_{1_{k_j}}}  =0  \Rightarrow \; \; \eta_{1_{k_j}}=\frac{1}{r_j}(\beta_{1_{k_j}}-\beta_{2_{k_j}}) 
\; ,  \ \text{or equivalently $\eta_{1_{k}}=R_H^{-1}(\beta_{1_{k}}-\beta_{2_{k}})$}\\
\label{necHeta2}
& \frac{\delta L}{\delta \eta_{2_{k}}}  =0  \Rightarrow
 \gamma_{1_{k}}+\gamma_{2_{k}}
=\beta_{1_{k}}-\beta_{2_{k}}-\varphi_{k} \\
\label{necHs}
& \frac{\delta L}{\delta s_{k}}  =0  \Rightarrow
 \beta_{2_{k}}+\beta_{1_{k}}=\kappa \; \; \text{where $\kappa$ is defined in (\ref{RHepskapdef})} \\
\label{necHx}
&    \frac{\delta L}{\delta \hat{x}_k}  =0 \text{ for 1$\leq$ k $\leq$ N} \Rightarrow  \lambda_{k-1} = A'\lambda_k + C'(\gamma_{1_k}-\gamma_{2_k})  \\
    \label{necHinitcondn}
&     \frac{\delta L}{\delta \hat{x}_0} =0 \Rightarrow \hat{x}_0 =\bar{x}_0 +P^{-1}A'\lambda_0  \\
    \label{necHlambdabound}
&    \frac{\delta L}{\delta \hat{x}_{N+1}}  =0 \Rightarrow  \lambda_N=0  \\
    \label{weq1H}
&    \frac{\delta L}{\delta \hat{w}_k}  =0  \Rightarrow  \hat{w}_k=Q^{-1}B'\lambda_k  \\
    \label{xeqH}
&     \frac{\delta L}{\delta \lambda_k} =0 \Rightarrow
    \hat{x}_{k+1} =  A\hat{x}_k+B\hat{w}_k \\
    \label{ineqconstrH}
 &   \gamma_{1_k}, \gamma_{2_k}. \beta_{1_k}, \beta_{2_k}, \varphi_k \geq 0 \; \; \text{due to the inequality constraints} \\
& \gamma_{1_k}'(\epsilon+\eta_{2_k}-y_k+C\hat{x}_k)=0 \ , \ \gamma_{2_k}'(\epsilon+\eta_{2_k}+y_k-C\hat{x}_k)=0 \ , \
\beta_{1_k}'(s_k+\eta_{1_k}-\eta_{2_k})=0 \ , \
\beta_{2_k}'(s_k-\eta_{1_k}+\eta_{2_k})=0 \ , \
\varphi_k' \eta_{2_k}=0
\end{align}
\end{subequations}
Substituting the above,
\begin{align}
    L & = \frac{1}{2} \left[ \lambda_0'AP^{-1}A'\lambda_0 
    + \sum_{k=0}^{N-1} \lambda_k'B'Q^{-1}B\lambda_k
    + \sum_{j=1}^{m} \sum_{k=1}^N \frac{1}{r_j} (\beta_{1_{k_j}}-\beta_{2_{k_j}})^2 \right]
    -\varepsilon'\sum_{k=1}^N (\gamma_{1_k}+\gamma_{2_k}) \nonumber \\
    & - \sum_{j=1}^{m} \sum_{k=1}^N \frac{1}{r_j} (\beta_{1_{k_j}}-\beta_{2_{k_j}})^2 + \sum_{k=1}^N (\gamma_{1_k}-\gamma_{_k})'(y_k-C\hat{x}_k)
\end{align}
As in the proof of Lemma \ref{Huberreg} one can show that a) $\gamma_{1_k}+\gamma_{2_k}=|\gamma_{1_k}-\gamma_{2_k}|=
\beta_{1_{k}}-\beta_{2_{k}}$ and b) $-\kappa \leq \gamma_{1_k}-\gamma_{2_k} \leq \kappa$. Let $\theta$ and $\Theta$ be defined as 
\[
\theta_k:=\gamma_{1_k}-\gamma_{2_k} \; , \Theta:=\begin{bmatrix} \theta_1 \\ \vdots \\ \theta_N \end{bmatrix}
\]
One can proceed as in the proof of \ref{theorem1} and the optimality conditions above to show that $\lambda_0=\begin{bmatrix} C' & A'C' & \cdots & A'^{N-1}C' \end{bmatrix} \Theta$ and 
\begin{align}
L & =\frac{1}{2} \left[  \lambda_0' A P^{-1} A' \lambda_0 + \Theta'FQ_{inv}F'\Theta + \sum_{j=1}^{m} \sum_{k=1}^N \frac{1}{r_j} \theta_{k_j}^2 \right] 
- \epsilon' \sum_{k=1}^{N} |\theta_k| - \sum_{j=1}^{m} \sum_{k=1}^N \frac{1}{r_j} \theta_{k_j}^2  - \Theta' \begin{bmatrix} C\hat{x}_1 \\ \vdots \\ C\hat{x}_N \end{bmatrix} + \Theta' \begin{bmatrix} y_1 \\ \vdots \\ y_N \end{bmatrix} \nonumber \\
& =\frac{1}{2} \left[  \lambda_0' A P^{-1} A' \lambda_0 + \Theta'FQ_{inv}F'\Theta +  \sum_{k=1}^N  \theta_{k}'R_H^{-1} \theta_{k} \right] 
- \epsilon' \sum_{k=1}^{N} |\theta_k| - \sum_{k=1}^N   \theta_{k}'R_H^{-1} \theta_{k}  - \Theta' \begin{bmatrix} C\hat{x}_1 \\ \vdots \\ C\hat{x}_N \end{bmatrix} + \Theta' \begin{bmatrix} y_1 \\ \vdots \\ y_N \end{bmatrix} \nonumber \\
& = - \frac{1}{2}  \Theta' M_H \Theta - \epsilon' \sum_{k=1}^{N} |\theta_k|  + \Theta' Y \; \text{ where $M_H$ and $Y$ are as defined in (\ref{MHdef}) and (\ref{FGdef}) } \nonumber
\end{align}
Rest of the proof follows as in the proof of Theorem \ref{theorem1} except that there is an additional constraint that $|\theta_k| \leq \kappa$.

\subsection{Proof of Theorem \ref{theorem3}}
The optimization problem (\ref{problem2}) subject to constraints (\ref{constrdynamics}), (\ref{constr2b}) and (\ref{constr3c}) can be written in terms of its Lagrangian as follows where $\lambda_k \in \mathbb{R}^n$, $\gamma_k$ and $\beta_k \in \mathbb{R}^m$ and $\xi \in \mathbb{R}^l$ are the Lagrange multipliers:
\begin{align}
 & \max_{\lambda_k,\gamma_k, \beta_k, \xi} \text{   } \min_{\hat{x}_0,\hat{x}_k, \hat{w}_k} \;  L \;  \; \; \; \; \; \text{         with  } \gamma_k \geq 0  \; \beta_k \geq 0 \text{ and } \xi \geq 0   \nonumber \\
 \label{lagrangpr2}
\text{   where  } \; & L  =  \frac{1}{2} \left[ (\hat{x}_0-\bar{x}_0)'P(\hat{x}_0-\bar{x}_0) + \sum_{k=0}^{N-1} \hat{w}_k'Q \hat{w}_k + \sum_{k=1}^{N} (y_k-C\hat{x}_k-\eta_k)'R(y_k-C\hat{x}_k-\eta_k)  \right]  \nonumber \\
& + \sum_{k=0}^{N} \lambda_k' \left(\hat{x}_{k+1} -  A\hat{x}_k-B\hat{w}_k  \right)  - \sum_{k=1}^{N} \gamma_k'(\epsilon-\eta_k) - \sum_{k=1}^{N} \beta_k'(\epsilon+\eta_k) - \xi' \left( a- \sum_{k=1}^NU_k\hat{x}_k -   \sum_{k=0}^{N-1}V_k\hat{w}_k  \right)   
\end{align}

Similar to the proof above, the optimality conditions are 

\begin{subequations}
\begin{align}
\frac{\delta L}{\delta \hat{x}_k} & =0 \text{ for 1$\leq$ k $\leq$ N} \Rightarrow C'RC\hat{x}_k-C'R(y_k-\eta_k)+\lambda_{k-1}-A'\lambda_k + U_k'\xi=0 \nonumber \\
    \label{lambdaeq2}
     \;  \Rightarrow & \; \lambda_{k-1} = A'\lambda_k - C'RC \hat{x}_k + C'R(y_k-\eta_k) - U_k'\xi  \\
    \frac{\delta L}{\delta \hat{x}_0} & =0 \Rightarrow P\hat{x}_0 - P\bar{x}_0 - A' \lambda_0=0  
    \Rightarrow \hat{x}_0 =\bar{x}_0 +P^{-1}A'\lambda_0  \\
    \label{lambdabound2}
    \frac{\delta L}{\delta \hat{x}_{N+1}} & =0 \Rightarrow  \lambda_N=0  \\
    \label{weq2}
    \frac{\delta L}{\delta \hat{w}_k} & =0  \Rightarrow  Q\hat{w}_k - B'\lambda_k + V_k'\xi =0 \Rightarrow \hat{w}_k=Q^{-1}B'\lambda_k  - Q^{-1} V_k'\xi \\
    \label{etaeq2}
    \frac{\delta L}{\delta \eta_k} & =0  \Rightarrow  R\eta_k - R(y_k-C\hat{x}_k)+(\gamma_k-\beta_k)=0 
    \Rightarrow \eta_k = (y_k-C\hat{x}_k) -R^{-1}(\gamma_k-\beta_k)  \\
    \gamma_k & \geq 0, \text{  } \beta_k \geq 0, \xi \geq 0 \ , \ \gamma_k'(\epsilon-\eta_k)=0 \ , \beta_k'(\epsilon+\eta_k)=0 \ , \ \xi' \left( a- \sum_{k=1}^NU_k\hat{x}_k -   \sum_{k=0}^{N-1}V_k\hat{w}_k  \right)=0  \\
    \label{xeq2}
     \frac{\delta L}{\delta \lambda_k} & =0 \Rightarrow
    \hat{x}_{k+1} =  A\hat{x}_k+B\hat{w}_k
\end{align}
\end{subequations}

From (\ref{lambdaeq2}) to (\ref{xeq2}) and noting the definition of $\theta$ in (\ref{thetadef}) one observes that 

\begin{align}
    \label{lamdadynamics2}
    \lambda_{k-1}  & = A'\lambda_{k}+C'\theta_k -U_k' \xi \text{ , } \lambda_N=0 \Rightarrow \begin{bmatrix} B'\lambda_0 \\ \vdots \\ B'\lambda_{N-1} \end{bmatrix}= F' \Theta -G \xi \text{   where  F and G are defined in } (\ref{FGdef}) \text{ and } (\ref{VGHdef})  \\
    & \sum_{k=0}^{N-1} \hat{w}_k'Q \hat{w}_k =(\Theta' F -\xi' G'-\xi' V)Q_{inv}(F' \Theta -G \xi-V' \xi)  \text{   where V is defined in } (\ref{VGHdef})
\end{align}
\begin{align}
    \hat{x}_{k+1}  & =  A\hat{x}_k+BQ^{-1}B'\lambda_k - BQ^{-1} V_k' \xi \; , \; \hat{x}_0= \bar{x}_0 +P^{-1}A'\lambda_0
    \nonumber \\ 
    \label{Chatx2}
    & \Rightarrow 
\Theta' \begin{bmatrix} C\hat{x}_1 \\ \vdots \\ C\hat{x}_N \end{bmatrix} = \Theta' F Q_{inv} (F' \Theta - G \xi - V' \xi) 
+\Theta'\begin{bmatrix} CA \\ \vdots \\ CA^N \end{bmatrix} (\bar{x}_0 +P^{-1}A'\lambda_0) 
\end{align}

\begin{equation}
 \label{Whatw}
 \xi' \begin{bmatrix} V_0 & \cdots & V_{N-1} \end{bmatrix} \begin{bmatrix} \hat{w}_0 \\ \vdots \\ \hat{w}_{N-1} \end{bmatrix} 
     = \xi' VQ_{inv}(F' \Theta - G \xi - V' \xi)  
\end{equation}

\begin{equation}
 \label{Uhatx}
 \xi' \begin{bmatrix} U_1 & \cdots & U_N \end{bmatrix} \begin{bmatrix} \hat{x}_1 \\ \vdots \\ \hat{x}_N \end{bmatrix} 
     = \xi' G'Q_{inv}(F' \Theta - G \xi - V' \xi) +  \xi' ( \sum_{i=1}^{N} U_iA^i ) (\bar{x}_0 +P^{-1}A'\lambda_0) 
\end{equation}
From (\ref{lamdadynamics2}) one also observes that $\lambda_0=[C' \; A'C' \; \cdots \; A'^{N-1}C']\Theta-[\sum_{i=1}^{N}A'^{(i-1)}U_i']\xi$. This together with the above implies that
\begin{subequations}
\begin{align}
& \lambda_0'AP^{-1}A'\lambda_0  = \begin{bmatrix} \Theta & \xi \end{bmatrix}' H  P^{-1}
     H' \begin{bmatrix} \Theta \\ \xi \end{bmatrix} \; \text{where $H$ is as defined in (\ref{VGHdef})}  \\
& -\Theta'\begin{bmatrix} CA \\ \vdots \\ CA^N \end{bmatrix}  P^{-1}A'\lambda_0 +  \xi' ( \sum_{i=1}^{N} U_iA^i )  P^{-1}A'\lambda_0
     = - \begin{bmatrix} \Theta & \xi \end{bmatrix}' H  P^{-1}
     H' \begin{bmatrix} \Theta \\ \xi \end{bmatrix}  \\
& - \Theta' \begin{bmatrix} C\hat{x}_1 \\ \vdots \\ C\hat{x}_N \end{bmatrix} +  \xi' \begin{bmatrix} V_0 & \cdots & V_{N-1} \end{bmatrix} \begin{bmatrix} \hat{w}_0 \\ \vdots \\ \hat{w}_{N-1} \end{bmatrix} + \xi' \begin{bmatrix} U_1 & \cdots & U_N \end{bmatrix} \begin{bmatrix} \hat{x}_1 \\ \vdots \\ \hat{x}_N \end{bmatrix}  \nonumber \\
& = - (\Theta' F -\xi' G'-\xi' V)Q_{inv}(F' \Theta -G \xi-V' \xi) -\begin{bmatrix} \Theta & \xi \end{bmatrix}' H  P^{-1}
     H' \begin{bmatrix} \Theta \\ \xi \end{bmatrix} + \xi' (\sum_{i=1}^NU_iA^i )\bar{x}_0-\Theta'\begin{bmatrix} CA \\ \vdots \\ CA^N \end{bmatrix}\bar{x}_0
\end{align}
\end{subequations}

Using the above identities, the expression of Lagrangian (\ref{lagrangpr2}) simplifies into the following form after some algebraic simplification

\begin{align}
L  & =  \frac{1}{2} \left[\lambda_0'AP^{-1}A'\lambda_0 +  \sum_{k=0}^{N-1} \hat{w}_k'Q \hat{w}_k +\Theta'R_{inv}\Theta \right] 
-  \epsilon' \sum_{k=1}^{N} (\gamma_k+\beta_k) + \Theta'\begin{bmatrix} y_1 \\ \vdots \\ y_N \end{bmatrix} - \Theta' \begin{bmatrix} C\hat{x}_1 \\ \vdots \\ C\hat{x}_N \end{bmatrix}-\Theta'R_{inv}\Theta  \nonumber \\
 & \; \; \; \; + \xi' \begin{bmatrix} U_1 & \cdots & U_N \end{bmatrix} \begin{bmatrix} \hat{x}_1 \\ \vdots \\ \hat{x}_N \end{bmatrix} 
 +\xi' \begin{bmatrix} V_0 & \cdots & V_{N-1} \end{bmatrix} \begin{bmatrix} \hat{w}_0 \\ \vdots \\ \hat{w}_{N-1} \end{bmatrix}  
 - \xi'a \nonumber \\
& = - \frac{1}{2} \left[ \begin{bmatrix} \Theta & \xi \end{bmatrix}' H  P^{-1}
     H' \begin{bmatrix} \Theta \\ \xi \end{bmatrix}  +(\Theta' F -\xi' G'-\xi' V)Q_{inv}(F' \Theta -G \xi-V' \xi)+ \Theta'R_{inv}\Theta \right] \nonumber \\
      & \; \; \; \; \; -  \epsilon' \sum_{k=1}^{N} (\gamma_k+\beta_k) + \Theta' Y - \xi'(a-\sum_{i=1}^NU_iA^i\bar{x}_0) \nonumber \\
      \label{lagrangepsconstr}
      & = - \frac{1}{2} \begin{bmatrix} \Theta & \xi \end{bmatrix}' T \begin{bmatrix} \Theta \\ \xi \end{bmatrix} -  \epsilon' \sum_{k=1}^{N} (\gamma_k+\beta_k) + \Theta' Y - \xi'(a-\sum_{i=1}^NU_iA^i\bar{x}_0) \; \text{ where $T$ is defined in } (\ref{Tdef})
\end{align}

As in proof of the Theorem \ref{theorem1}, from KKT optimality conditions one notes that since $\epsilon>0$, if $\gamma_{k_i} >0$, then $\beta_{k_i}=0$ and if $\beta_{k_i}>0$ then $\gamma_{k_i}=0$. Thus just as in equation (\ref{thetamax}) here also for the optimal solution it is true that  $\gamma_k+\beta_k=|\gamma_k-\beta_k|=|\theta_k|$. Since $\epsilon>0$, as in the previous case one observes that for any $\Theta$ 
\[
-  \epsilon' \sum_{k=1}^{N} (\gamma_k+\beta_k) = -  \epsilon' \sum_{k=1}^{N} |\theta_k| = - \max_{\zeta} \; \varepsilon' \zeta \; \; \text{subject to constraints } \zeta \geq \Theta, \; \zeta \geq -\Theta
\]
The optimization problem described in (\ref{opteps2}) and (\ref{constreps2}) to obtain optimal $\Theta$ and $\xi$ follows from (\ref{lagrangepsconstr}) and the above.

\subsection{Proof of Theorem \ref{theorem5}}
The proof is similar to that of Theorem \ref{theorem3} and so we will only highlight key differences. The Lagrangian for the optimization problem (\ref{problempred}) subject to constraints (\ref{constrdynamics}), (\ref{constr2b}) and (\ref{constrpred}) can be described as follows where the main differences compared to the Lagrangian for Theorem \ref{theorem2}) are in upper limits of the summation in some of the terms :
\begin{align}
 & \max_{\lambda_k,\gamma_k, \beta_k, \xi} \text{   } \min_{\hat{x}_0,\hat{x}_k, \hat{w}_k} \;  L \;  \; \; \; \; \; \text{         with  } \gamma_k \geq 0  \; \beta_k \geq 0 \text{ and } \xi \geq 0   \nonumber \\
 \label{lagrangpr3}
\text{   where  } \; & L  =  \frac{1}{2} \left[ (\hat{x}_0-\bar{x}_0)'P(\hat{x}_0-\bar{x}_0) + \sum_{k=0}^{N+j-1} \hat{w}_k'Q \hat{w}_k + \sum_{k=1}^{N} (y_k-C\hat{x}_k-\eta_k)'R(y_k-C\hat{x}_k-\eta_k)  \right]  \nonumber \\
& + \sum_{k=0}^{N+j} \lambda_k' \left(\hat{x}_{k+1} -  A\hat{x}_k-B\hat{w}_k  \right)  - \sum_{k=1}^{N} \gamma_k'(\epsilon-\eta_k) - \sum_{k=1}^{N} \beta_k'(\epsilon+\eta_k) - \xi'\left( a- \sum_{k=1}^{N+j}U_k\hat{x}_k -   \sum_{k=0}^{N+j-1}V_k\hat{w}_k  \right)   
\end{align}
As in the previous case, $\lambda_k \in \mathbb{R}^n$, $\gamma_k$ and $\beta_k \in \mathbb{R}^m$ and $\xi \in \mathbb{R}^l$ are the Lagrange multipliers with $\gamma_k \geq 0$, $\beta_k \geq 0$ and $\xi \geq 0$ due to the inequality constraints. The necessary conditions for optimality are 

\begin{subequations}
\begin{align}
\frac{\delta L}{\delta \hat{x}_k} & =0 \text{ for 1$\leq$ k $\leq$ N} \Rightarrow C'RC\hat{x}_k-C'R(y_k-\eta_k)+\lambda_{k-1}-A'\lambda_k + U_k'\xi=0 \nonumber \\
    \label{lambdaeq3a}
     \;  \Rightarrow & \; \lambda_{k-1} = A'\lambda_k - C'RC \hat{x}_k + C'R(y_k-\eta_k) - U_k'\xi  \; \; \text{ for 1$\leq$ k $\leq$ N} \\
     \label{lambdaeq3b}
     \frac{\delta L}{\delta \hat{x}_k} & =0 \text{ for N+1$\leq$ k $\leq$ N+j } \Rightarrow \lambda_{k-1} = A'\lambda_k - U_k'\xi \; \;  \text{ for N+1$\leq$ k $\leq$ N+j}\\
    \frac{\delta L}{\delta \hat{x}_0} & =0 \Rightarrow P\hat{x}_0 - P\bar{x}_0 - A' \lambda_0=0  
    \Rightarrow \hat{x}_0 =\bar{x}_0 +P^{-1}A'\lambda_0  \\
    \label{lambdabound3}
    \frac{\delta L}{\delta \hat{x}_{N+j+1}} & =0 \Rightarrow  \lambda_{N+j}=0  \\
    \label{weq3}
    \frac{\delta L}{\delta \hat{w}_k} & =0  \Rightarrow  Q\hat{w}_k - B'\lambda_k + V_k'\xi =0 \Rightarrow \hat{w}_k=Q^{-1}B'\lambda_k  - Q^{-1} V_k'\xi \\
    \label{etaeq3}
    \frac{\delta L}{\delta \eta_k} & =0  \Rightarrow  R\eta_k - R(y_k-C\hat{x}_k)+(\gamma_k-\beta_k)=0 
    \Rightarrow \eta_k = (y_k-C\hat{x}_k) -R^{-1}(\gamma_k-\beta_k)  \\
    \gamma_k & \geq 0, \text{  } \beta_k \geq 0, \xi \geq 0 \; , \ 
    \gamma_k'(\epsilon-\eta_k)=0 \ , \ \beta_k'(\epsilon+\eta_k)=0 \ , \ \xi'\left( a- \sum_{k=1}^{N+j}U_k\hat{x}_k -   \sum_{k=0}^{N+j-1}V_k\hat{w}_k  \right)=0 \\ 
    \label{xeq3}
     \frac{\delta L}{\delta \lambda_k} & =0 \Rightarrow
    \hat{x}_{k+1} =  A\hat{x}_k+B\hat{w}_k
\end{align}
\end{subequations}

Equations (\ref{optlambdaep3a}) and (\ref{optlambdaep3b}) for obtaining optimal $\lambda_k$ follow from (\ref{lambdaeq3a}), (\ref{lambdaeq3b}), (\ref{lambdabound3}) and (\ref{etaeq3}). From (\ref{optlambdaep3a}) and (\ref{optlambdaep3b}), one also notes 
\begin{subequations}
\begin{align}
\begin{bmatrix} B'\lambda_0 \\ \vdots \\ B'\lambda_{N+j-1} \end{bmatrix} & = \bar{F}' \Theta - \bar{G} \xi \;,  \; \;
\text{where $\bar{F}$ and $\bar{G}$ are defined in (\ref{FGdefa}) and (\ref{FGdefb})}\\
\sum_{k=0}^{N+j-1} \hat{w}_k'Q \hat{w}_k & =(\Theta' \bar{F} -\xi' \bar{G}'-\xi' \bar{V})\bar{Q}_{inv}(\bar{F}' \Theta -\bar{G} \xi-\bar{V}' \xi) \; , \;  \text{   where $\bar{V}$ is defined in } (\ref{FGdefb})
\end{align}
\end{subequations}
From equations (\ref{optlambdaep3a}) and (\ref{optlambdaep3b}) one observes that $\lambda_0=[C' \; A'C' \; \cdots \; A'^{N-1}C']\Theta-[\sum_{i=1}^{N+j}A'^{(i-1)}U_i']\xi$. As in the proof of the previous Theorem, one can show that
\[
\lambda_0'AP^{-1}A'\lambda_0 = \begin{bmatrix} \Theta & \xi \end{bmatrix}' \bar{H}  P^{-1}
     \bar{H}' \begin{bmatrix} \Theta \\ \xi \end{bmatrix} \; \; \text{where $\bar{H}$ is as defined in (\ref{FGdefb})}
\]
\begin{subequations}
\begin{align}
& - \Theta' \begin{bmatrix} C\hat{x}_1 \\ \vdots \\ C\hat{x}_N \end{bmatrix} +  \xi' \begin{bmatrix} V_0 & \cdots & V_{N+j-1} \end{bmatrix} \begin{bmatrix} \hat{w}_0 \\ \vdots \\ \hat{w}_{N+j-1} \end{bmatrix} + \xi' \begin{bmatrix} U_1 & \cdots & U_N \end{bmatrix} \begin{bmatrix} \hat{x}_1 \\ \vdots \\ \hat{x}_{N+j} \end{bmatrix}  \nonumber \\
& = - (\Theta' \bar{F} -\xi' \bar{G}'-\xi' \bar{V})\bar{Q}_{inv}(\bar{F}' \Theta -\bar{G} \xi-\bar{V}' \xi) -\begin{bmatrix} \Theta & \xi \end{bmatrix}' \bar{H}  P^{-1}
     \bar{H}' \begin{bmatrix} \Theta \\ \xi \end{bmatrix} + \xi' (\sum_{i=1}^{N+j}U_iA^i )\bar{x}_0-\Theta'\begin{bmatrix} CA \\ \vdots \\ CA^N \end{bmatrix}\bar{x}_0 \nonumber
\end{align}
\end{subequations}
Using the above, the expression of Lagrangian (\ref{lagrangpr3}) simplifies into the following form after some algebraic simplification
\begin{align}
L  & =  \frac{1}{2} \left[\lambda_0'AP^{-1}A'\lambda_0 +  \sum_{k=0}^{N+j-1} \hat{w}_k'Q \hat{w}_k +\Theta'R_{inv}\Theta \right] 
-  \epsilon' \sum_{k=1}^{N} (\gamma_k+\beta_k) + \Theta'\begin{bmatrix} y_1 \\ \vdots \\ y_N \end{bmatrix} - \Theta' \begin{bmatrix} C\hat{x}_1 \\ \vdots \\ C\hat{x}_N \end{bmatrix}-\Theta'R_{inv}\Theta  \nonumber \\
 & \; \; \; \; + \xi' \begin{bmatrix} U_1 & \cdots & U_{N+j} \end{bmatrix} \begin{bmatrix} \hat{x}_1 \\ \vdots \\ \hat{x}_{N+j} \end{bmatrix} 
 +\xi' \begin{bmatrix} V_0 & \cdots & V_{N+j-1} \end{bmatrix} \begin{bmatrix} \hat{w}_0 \\ \vdots \\ \hat{w}_{N+j-1} \end{bmatrix}  
 - \xi'a \nonumber \\
& = - \frac{1}{2} \left[ \begin{bmatrix} \Theta & \xi \end{bmatrix}' \bar{H}  P^{-1}
     \bar{H}' \begin{bmatrix} \Theta \\ \xi \end{bmatrix}  +(\Theta' \bar{F} -\xi' \bar{G}'-\xi' \bar{V})Q_{inv}(\bar{F}' \Theta -\bar{G} \xi-\bar{V}' \xi)+ \Theta'R_{inv}\Theta \right] \nonumber \\
      & \; \; \; \; \; -  \epsilon' \sum_{k=1}^{N} (\gamma_k+\beta_k) + \Theta' Y - \xi'(a-\sum_{i=1}^{N+j}U_iA^i\bar{x}_0) \nonumber \\
      \label{lagrang2}
      & = - \frac{1}{2} \begin{bmatrix} \Theta & \xi \end{bmatrix}' \bar{T} \begin{bmatrix} \Theta \\ \xi \end{bmatrix} -  \epsilon' \sum_{k=1}^{N} (\gamma_k+\beta_k) + \Theta' Y - \xi'(a-\sum_{i=1}^{N+j}U_iA^i\bar{x}_0) \; \text{ where $\bar{T}$ is defined in } (\ref{Tdefa}) \nonumber
\end{align}
As before, KKT optimality condition implies that since $\epsilon >0$, $\gamma_k+\beta_k=|\gamma_k-\beta_k|=|\theta_k|$. Rest of the proof follows along the same lines as the one in the previous section.

\subsection{Proof of Theorems \ref{theorem4} and \ref{theorem6}}

Proof of Theorems \ref{theorem4} and \ref{theorem6} are analogous to proofs of Theorems \ref{theorem3} and \ref{theorem5} where the Huber cost function is incorporated as in (\ref{Hubercostproof}) (within the proof of Theorem \ref{theorem2}).

\bibliographystyle{unsrt}  

\end{document}